\documentclass[3p,onecolumn,times]{elsarticle}  % Elsevier journal class

\makeatletter
\def\ps@pprintTitle{%
 \let\@oddhead\@empty
 \let\@evenhead\@empty
 \def\@oddfoot{\reset@font\hfil} % Remove preprint footnote
 \let\@evenfoot\@oddfoot}
\makeatother

\usepackage{geometry}
\usepackage{graphicx}               % For including figures
\usepackage{dcolumn}                % Align table columns on decimal point
\usepackage{bm}                     % Bold math
\usepackage{hyperref}               % Add hyperlinks
\usepackage{xcolor}                 % Colors for hyperlinks, tables, etc.
\hypersetup{
    colorlinks=true,                % Enable colored links
    linkcolor=blue,                 % Color of internal links
    citecolor=red,                 % Color of citation links
    filecolor=magenta,              % Color of file links
    urlcolor=cyan                   % Color of external links
}
\usepackage{amsmath}                % Advanced math typesetting
\usepackage{amsfonts}               % For special fonts
\usepackage{amssymb}                % For more special symbols
\usepackage{siunitx}                % For SI units
\usepackage{float}                  % For precise figure and table placement
\usepackage{comment}                % For block comments
\usepackage{enumitem}               % Customizable list environments
\usepackage{mathtools}              % Advanced math formatting
\usepackage{physics}                % Physics-related symbols
\usepackage{setspace}               % Control line spacing
\usepackage{datetime2}              % Custom date format
\usepackage{multirow}               % For multi-row tables
\usepackage{booktabs}               % Professional-looking tables
\usepackage{microtype}              % Better typography
\usepackage{tikz}                   % Drawing and plots
\usepackage{pgfplots}               % For creating plots
\usepackage{caption}
\usepackage{subcaption}
\usetikzlibrary{decorations.pathmorphing}
\DeclareUnicodeCharacter{0301}{\'{e}}
\pgfplotsset{compat=1.18}
\usepackage{url} % To format URLs
\usepackage[utf8]{inputenc}
\AtBeginDocument{\RenewCommandCopy\qty\SI}
\DeclareUnicodeCharacter{2212}{\textminus}
\DeclareUnicodeCharacter{039B}{$\Lambda$}
\DeclareUnicodeCharacter{2131}{\mathcal{F}}
\DeclareUnicodeCharacter{03A9}{$\Omega$}
\DeclareUnicodeCharacter{223C}{\sim}
\DeclareUnicodeCharacter{03B1}{$\alpha$}
\usepackage[T1]{fontenc}
\usepackage{lmodern}                % Modern font package

\begin{document}

\begin{frontmatter}

\title{\Large Reconstructions of Einstein-Aether Gravity from Barrow Agegraphic and New Barrow Agegraphic Dark Energy models: Examinations and Observational Limits}

\author[1]{Banadipa Chakraborty~\href{https://orcid.org/0009-0008-2378-0544}{\includegraphics[height=1em]{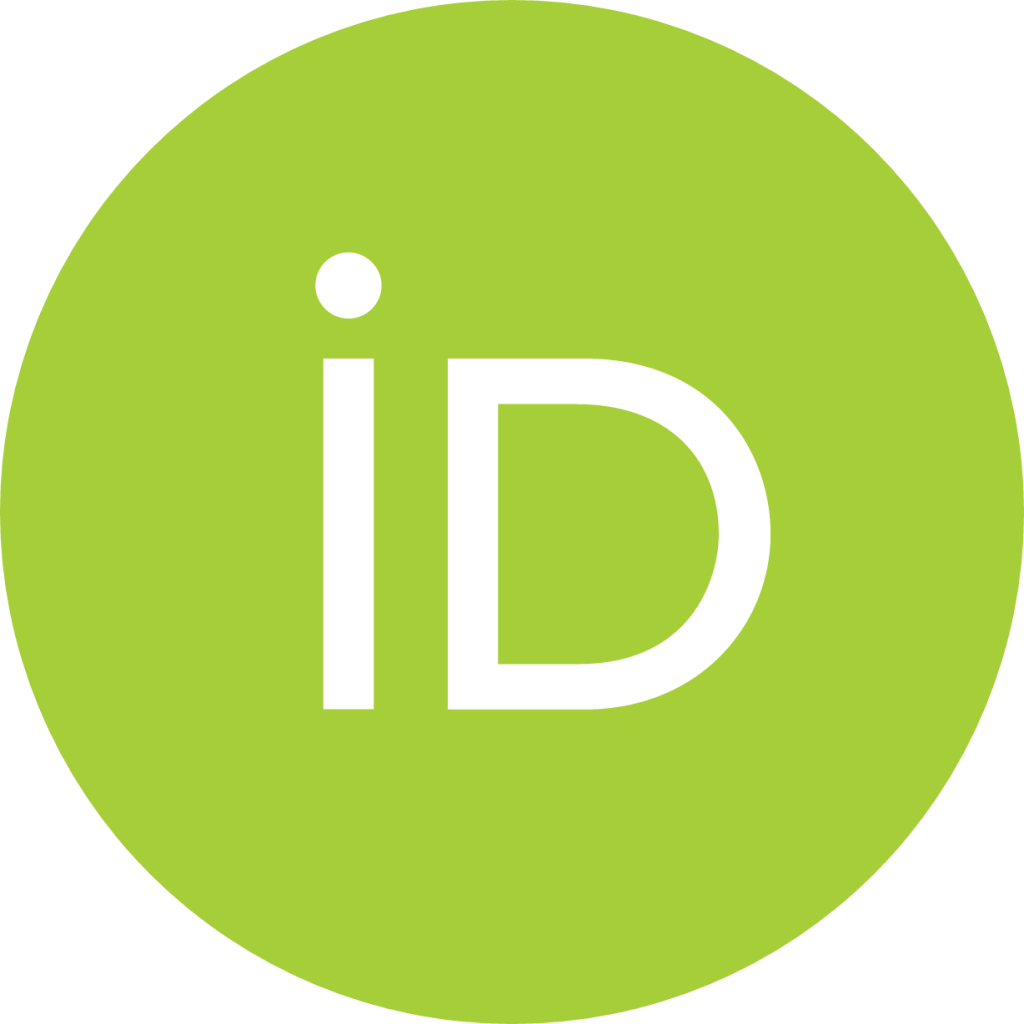}}}
\ead{jhelum.chakraborty@gmail.com}

\author[1]{Tamal Mukhopadhyay~\href{https://orcid.org/0000-0001-9843-906X}{\includegraphics[height=1em]{ORCID_iD_svg.png}}}
\ead{tamalmukhopadhyay7@gmail.com}

\author[2]{Anamika Kotal~\href{https://orcid.org/0009-0009-8426-3206}{\includegraphics[height=1em]{ORCID_iD_svg.png}}}
\ead{kotalanamika31@gmail.com}

\author[2]{Ujjal Debnath~\href{https://orcid.org/0000-0002-2124-8908}{\includegraphics[height=1em]{ORCID_iD_svg.png}}}
\ead{ujjaldebnath@gmail.com}

\address[1]{Department of Physics, Sister Nivedita University, DG Block (Newtown) 1/2, Action Area I, Kolkata-700156, India}
\address[2]{Department of Mathematics, Indian Institute of Engineering Science and Technology, Shibpur, Howrah-711103, India}

\begin{abstract}
We present a comprehensive investigation exploring the theoretical framework of Einstein-Aether gravity theory when combined with two modified cosmological paradigms: the Barrow Agegraphic Dark Energy (BADE) and its newer variant, the New Barrow Agegraphic Dark Energy (NBADE). Our study focuses on reconstructing the functional form of the Einstein-Aether Lagrangian component $F(K)$ from these phenomenological dark energy models. Model parameters are constrained using a Markov Chain Monte Carlo (MCMC) approach based on multiple datasets, including cosmic chronometers (CC), Baryon Acoustic Oscillations (BAO), and the Pantheon+SH0ES compilation. Using best-fit parameters, we analyze various cosmological diagnostics: Hubble and deceleration parameter evolution, dark energy equation of state ($\omega_{DE}$), density parameter trajectories, $\omega_{DE}'$–$\omega_{DE}$ phase space behavior, statefinder diagnostics $(r,s^*)$ and $(r,q)$, and Om(z) trajectories. Both models exhibit late-time acceleration, with the dark energy sector showing a quintessence-like nature in the current epoch and evolving toward a phantom regime in the future. Stability analysis based on the squared sound speed ($v_s^2$) highlights partial epoch-dependent stability. While our results demonstrate reasonable agreement with observational data and reveal physically plausible dynamics, the models do not yet offer a fundamentally superior alternative to other dark energy reconstructions. Nonetheless, their behavior under modified entropy assumptions and their flexibility in dynamical diagnostics provide a useful framework for probing non-standard extensions of Einstein-Aether gravity and dark energy phenomenology.
\end{abstract}


\end{frontmatter}

\section{Introduction}\label{Sect: Intro}
The latter part of the 20\textsuperscript{th} century witnessed a dramatic change in cosmology with the emergence of compelling observational evidence supporting the accelerated expansion of the universe. This phenomenon, which appears to have commenced relatively recently in cosmic history, has been corroborated by multiple independent lines of evidence \cite{shapiro2006we}. Notably, observations of Type Ia supernovae \cite{riess1998observational, Perlmutter_1998, perlmutter1999measurements}, analyses of large-scale cosmic structure \cite{colless20012df, PhysRevD.69.103501}, measurements of Cosmic Microwave Background Radiation (CMBR) anisotropies \cite{komatsu2009five, ade2014planck}, and studies of Baryon Acoustic Oscillations \cite{cole20052df, percival2007measuring} have all contributed to this paradigm-altering discovery. Despite the remarkable success of General Relativity in explaining numerous astrophysical phenomena, it fails to provide a satisfactory explanation for this late-time accelerated expansion. Despite its widespread acceptance, the prevailing $\Lambda$CDM cosmological paradigm, which employs a cosmological constant $\Lambda$ to account for dark energy phenomena, encounters significant theoretical challenges. Among these challenges, the ``Cosmological Coincidence Problem'' stands as a particularly perplexing issue that demands resolution \cite{velten2014aspects, yoo2012theoretical}. Moreover, a fundamental inconsistency exists between particle physics and cosmological observations: the well-established Standard Model of Particle Physics fails to provide a theoretical foundation for the Cold Dark Matter (CDM) particles that are fundamental to the $\Lambda$CDM framework \cite{weinberg2001cosmological}. These discrepancies raise important questions about our current understanding of the universe's composition and evolution.

Various dynamic dark energy models have been proposed to tackle these challenges. Typically, these models introduce an extra degree of freedom through a scalar field, whose Lagrangian includes either a potential or kinetic term, both of which can lead to accelerated cosmic expansion. The quintessence model \cite{peebles1988cosmology, Caldwell_1998, zlatev1999quintessence, carroll1998quintessence} has appeared as a particularly favored candidate for dark energy among physicists. Additionally, alternative dark energy models featuring dynamic vacuum energy have been explored, including tachyon fields \cite{Sen_2002}, k-essence \cite{armendariz2001essentials}, H-essence \cite{Wei_2005}, dilaton \cite{Gasperini_2001}, phantom fields \cite{Caldwell_2002}, Chaplygin gas \cite{GORINI_2006}, DBI-quintessence \cite{Gumjudpai_2009}, and DBI-essence \cite{martin2008dbi}.

A different approach to addressing the issue of cosmic acceleration is by altering the geometric component of the Einstein-Hilbert action. This modification can potentially account for the observed accelerated expansion without the necessity of introducing an ad hoc dark energy component (For detailed reviews, refer to \cite{clifton2012modified, nojiri2011unified, nojiri2017modified}). The idea of altering the standard theory of gravity was first investigated through $f(R)$ gravity, which extends the Einstein-Hilbert action by incorporating a function of the Ricci scalar curvature \cite{sotiriou2010f}. This pioneering work has inspired the development of numerous modified gravity models, including but not limited to those discussed in \cite{Rastkar_2011, Nojiri_2005, ferraro2011non, PhysRevD.84.024020, Jamil_2012, Bamba_2010, Myrzakulov_2012}. For a more extensive review  of modified gravitational models, refer to \cite{saridakis2021modified, nojiri2007introduction, clifton2012modified}. Einstein-Aether gravity represents a theoretically significant modified gravity model that introduces a unit, time-like vector field (the ``aether'') alongside the metric tensor in Einstein's equations \cite{jacobson2001gravity}. This approach is fundamentally motivated by considering potential Lorentz symmetry violations at high energies, which various quantum gravity approaches suggest might exist. The aether field establishes a preferred reference frame for the universe, effectively reintroducing an absolute time concept while preserving general covariance of the theory's mathematical structure. This characteristic makes Einstein-Aether theory particularly relevant for cosmological applications, as it naturally aligns with the cosmic time coordinate implicit in the FLRW metric. Several significant developments in Einstein-Aether gravity have emerged through studies investigating black hole solutions \cite{eling2006black}, cosmological perturbations \cite{saga2013generation}, axionic extensions \cite{balakin2016axionic}, and diverse cosmological applications \cite{pasqua2017cosmological, heinicke2005einstein}. Of particular relevance to our current study, Einstein-Aether theory provides a natural framework for accommodating time-dependent cosmological phenomena, making it an ideal candidate for integration with time-based dark energy models such as those derived from Barrow entropy considerations.

In the field of cosmology, the conventional approach involves solving field equations to derive the space-time geometry. However, the reconstruction method presents an alternative paradigm wherein the underlying geometry is presumed based on observational evidence or theoretical considerations, and subsequently, model equations consistent with empirical observations are derived \cite{PhysRevD.74.086005, nojiri2006dark, PhysRevD.77.106005, nojiri2024well, GADBAIL2022137509}.
Recent advancements in theoretical physics have sparked a growing interest in investigating possible connections between different Holographic Dark Energy (HDE) models and various modified gravity theories~\cite{Tamal_f(P)_f(Q),Ujjal_F(P),Pradhan_f(Q_T),Pameli_f(P)}. This field of research has garnered significant attention, as it presents promising prospects for shedding light on the mysterious nature of dark energy and its far-reaching implications for cosmology. The intersection of HDE models and modified gravity theories offers a rich area for exploration, with the capability to offer a deeper insight into the fundamental structure of the universe and the processes responsible for its accelerated expansion. By examining the interplay between these theoretical frameworks, researchers aim to construct more comprehensive and robust models that can reconcile observational data with theoretical predictions, thereby elevating our concept of the evolution of the universe and the fundamental aspects of dark energy. The evolution of Einstein-Aether gravity research has seen significant developments through various theoretical frameworks. A pioneering study was conducted examining both conventional and entropy-corrected versions of holographic and new agegraphic dark energy models within the Einstein-Aether gravity context \cite{debnath2014reconstructions}. This investigation provided crucial insights into the reconstructed function's behavior and its cosmological significance. Further theoretical advancement came through the analysis of Einstein-Aether gravity's reconstruction from diverse modified gravity theories, with particular attention to their cosmological stability characteristics \cite{ranjit2014reconstruction}. Subsequent research established meaningful connections between Einstein-Aether gravity and various scalar field dark energy models, revealing important temporal patterns in both scalar fields and their associated potentials throughout cosmic history \cite{debnath2015correspondence}. The theoretical framework was further expanded through investigations incorporating multiple holographic dark energy variants, including Tsallis, Renyi, and Sharma-Mittal formulations \cite{rani2019cosmological}. These studies utilized various analytical tools, such as the equation of state parameter $\omega_{DE}$ and the $\omega'_{DE}-\omega_{DE}$ phase plane, to explore their cosmological implications. The theoretical foundation for these holographic approaches originates from 't Hooft's groundbreaking work \cite{hooft1993dimensional}, which built upon fundamental principles of black hole thermodynamics \cite{bekenstein1973black, hawking1975particle}. This holographic principle proposes a fundamental relationship between volumetric space information and its boundary representation, conceptually similar to how two-dimensional holograms capture three-dimensional information. Many researchers now consider the Holographic principle fundamental in the quantum description of gravity. Applications of Holographic dark energy models have been explored extensively \cite{li2004model,li2009probing,myung2011entropic,nojiri2006unifying,nojiri2019holographic,nojiri2020unifying}. Recently, significant attention has been drawn to the newly introduced Barrow entropy correction for black holes \cite{Barrow_2020}. Barrow Holographic Dark Energy (BHDE), as developed by Saridakis \cite{Saridakis_2020}, emerges from this corrected entropy and offers promising improvements over conventional holographic dark energy models \cite{anagnostopoulos2020observational,adhikary2021barrow,srivastava2021barrow}. The motivation for this study lies in the broader applicability of Barrow entropy, which provides a more generalized framework than the Bekenstein-Hawking entropy. Barrow entropy's resemblance to Tsallis non-extensive entropy \cite{TsallisEntropy, Wilk_2000, Tsallis_2013} further underscores its relevance, while in a specific limit, it recovers the standard Bekenstein-Hawking entropy. This quantum-gravitational deformation correction to the black hole horizon may offer a more accurate description of cosmological phenomena, potentially resolving long-standing puzzles in the field. As demonstrated in Ref.~\cite{PhysRevD.102.123525}, BHDE effectively captures the thermal evolution of the universe throughout the periods dominated by dark energy and dark matter. Moreover, recent work has shown that Barrow Entropic Dark Energy is a subset of a more general Holographic Dark Energy framework when an appropriate IR cutoff is chosen \cite{nojiri2022barrow}. A recent modification inspired by Barrow entropy was proposed in \cite{Sharma_2021}, where the infrared (IR) cutoff is selected as the universe's age and the conformal time, leading to the development of Barrow Agegraphic Dark Energy (BADE) and New Barrow Agegraphic Dark Energy (NBADE) models. The authors have shown several interesting cosmological features in their model. A particularly important aspect of our study is the comparative analysis between BADE and NBADE models. While both incorporate Barrow entropy corrections, they differ fundamentally in their choice of infrared cutoff: BADE employs the universe's age ($T$), while NBADE utilizes conformal time ($\eta$). This distinction is not merely a mathematical convenience but represents different physical interpretations of the relevant cosmological scale. The conformal time cutoff in NBADE offers several theoretical advantages over the conventional age parameter: it tracks the maximum comoving distance that light could have traveled since the Big Bang, thus providing a more complete accounting of the universe's causal structure. This choice also addresses a critical issue in the original agegraphic models—namely that models using cosmic age ($T$) as the IR cutoff encounter difficulties in reproducing the matter-dominated era \cite{wei2008new}. Furthermore, conformal time naturally connects to the particle horizon, which has stronger theoretical foundations in quantum gravitational approaches to cosmology. Our comparative reconstruction of Einstein-Aether gravity from both models allows us to evaluate which IR cutoff yields more physically consistent results when confronted with observational data, potentially revealing deeper insights into the nature of quantum gravity effects on cosmological scales. The synergy between Einstein-Aether gravity and Barrow Agegraphic Dark Energy models presents a particularly compelling theoretical framework for several reasons. Einstein-Aether theory's introduction of a preferred cosmic time direction through its timelike vector field naturally complements the time-based infrared cutoffs fundamental to Agegraphic dark energy models. This temporal alignment provides a consistent theoretical foundation for investigating cosmic evolution across different epochs. Moreover, both frameworks address different aspects of the same fundamental cosmological puzzles: while Einstein-Aether gravity modifies the geometric sector to account for potential Lorentz invariance violations at high energies, Barrow entropy introduces quantum gravitational corrections to conventional holographic principles. By reconstructing Einstein-Aether gravity within the context of Barrow Agegraphic models, we can explore how quantum-gravitational deformations of entropy might manifest within a modified gravitational framework that already accommodates a preferred cosmic time. This approach not only offers new perspectives on the accelerated expansion problem but also provides potential insights into the quantum nature of spacetime itself, possibly bridging phenomenological cosmology with more fundamental physical principles. Previous studies have explored various holographic dark energy models within Einstein-Aether framework \cite{debnath2014reconstructions, rani2019cosmological}, but the unique quantum-gravitational aspects of Barrow entropy and its implications for Einstein-Aether reconstruction remain unexplored territory with significant theoretical promise. The confluence of these theoretical and observational developments motivates our investigation into the reconstruction of the functional forms of Einstein-Aether gravity using BADE and NBADE as the background evolution. The selection of BADE and NBADE as frameworks for reconstructing Einstein-Aether gravity is motivated by several theoretical considerations. Unlike conventional dark energy models, these formulations incorporate quantum gravitational effects through Barrow's entropy corrections, which account for potential fractal structures in black hole horizons. This quantum-gravitational foundation aligns conceptually with the motivations behind Einstein-Aether theory, which similarly seeks to address potential fundamental modifications to spacetime structure. Furthermore, the age of the universe and conformal time, which serve as infrared cutoffs in BADE and NBADE respectively, provide natural cosmological scales that can be meaningfully interpreted within Einstein-Aether's preferred reference frame. The time-based nature of these cutoffs makes them particularly suitable for reconstruction within a framework that already incorporates a preferred cosmic time direction. This theoretical congruence suggests that exploring the reconstruction of Einstein-Aether gravity from these specific dark energy models may yield insights that would be inaccessible through other combinations of modified gravity and dark energy frameworks. Our research methodology incorporates observational data from multiple sources to limit the parameters of our proposed model:

\begin{itemize}
    \item 31 direct Hubble parameter measurements extracted from Cosmic Chronometers (CC) datasets
    \item 26 data points sourced from Baryon Acoustic Oscillation (BAO) observations
    \item 1701 data points from the Pantheon+SH0ES compilation
\end{itemize}

By leveraging this comprehensive empirical dataset, we aim to establish robust observational bounds on the model parameters, thereby facilitating a more profound understanding of the underlying physical scenarios. The novelty of our approach lies in the unique combination of Einstein-Aether gravity with Barrow entropy-based agegraphic models, creating a theoretical framework that simultaneously addresses potential Lorentz violations and quantum gravitational effects in cosmological evolution. To achieve this, we employ the covariance matrix technique, which enables us to minimize associated errors and determine optimal values for several key parameters within our proposed model. This multi-faceted approach, combining CC, BAO, and Pantheon+SH0ES data, provides a more comprehensive and statistically significant basis for our analysis than previous studies in this domain. The inclusion of the Pantheon+SH0ES dataset, with its extensive 1701 data points, significantly enhances the statistical power of our study, allowing for more precise constraints on the model parameters and potentially revealing new insights into the interplay between modified gravity and quantum-gravitationally motivated dark energy. Our methodology not only aims to reconstruct the functional forms of Einstein-Aether gravity from BADE and NBADE but also seeks to elucidate how these frameworks might collectively address fundamental questions about the nature of spacetime, cosmic acceleration, and the quantum structure of the universe. The organization of this paper is as follows: in Sect.~\ref{Sect: Cosmological Parameters} We have reviewed some important cosmological parameters in Sect.~\ref{Sect: EA Model} we have presented the Einstein-Aether gravity in Sect.~\ref{Sect: BADE Model} we have discussed in brief about the Barrow agegraphic dark energy in Sect.~\ref{Sect: Reconstruction BADE} and Sect.~\ref{Sect: Reconstruction NBADE} we have reconstructed Einstein-Aether gravity from BADE and NBADE. Sect.~\ref{Sect: Methodology} We have discussed the methodology used in our analysis, Sect.~\ref{Sect: Result} is dedicated to the findings of our analysis, and finally, we end our paper with concluding remarks in Sect.~\ref{Sect: Conclusion}.

\section{Review of Important Parameters}\label{Sect: Cosmological Parameters}
The cosmographic parameters are constructed by expanding the Hubble parameter as a Taylor series around $z=0$, the present epoch \cite{MattVisser_2004}:

\begin{equation}
\label{eq:cosmographic_parameters}
q \equiv -\frac{\ddot{a}}{H^2a}, \quad
j \equiv \frac{\dddot{a}}{H^3a}, \quad
s \equiv \frac{\ddddot{a}}{H^4a}, \quad
l \equiv \frac{a^{(5)}}{H^5a}
\end{equation}

Here, $q$, $j$, $s$, and $l$ represent the deceleration, jerk, snap, and lerk parameters, respectively. These parameters describe various aspects of the universe's expansion:

\begin{itemize}
    \item $q$: Describes the rate of deceleration or acceleration of cosmic expansion.
    \item $j$: Quantifies the change in the acceleration rate.
    \item $s$ and $l$: Further describe the evolution of the jerk and snap, respectively.
\end{itemize}

Recent research indicates that the jerk parameter could offer a distinct approach for probing the spatial curvature of the universe \cite{caldwell2004expansion}. For further discussion on cosmographic parameters in modified gravity models, see \cite{capozziello2019extended, CAPOZZIELLO2022137229, CAPOZZIELLO2023101346}.

The cosmographic parameters are related by the following equations:

\begin{align}
q &= -1-\frac{\dot{H}}{H^2} \\
j &= \frac{\ddot{H}}{H^3}-3q-2 = q(2q+1)+(1+z)\frac{dq}{dz} \\
s &= \frac{H^{(3)}}{H^4}+4j+3q(q+4)+6 = -(1+z)\frac{dj}{dz}-j(2+3q) \\
l &= \frac{H^{(4)}}{H^5}-24-6q-3q^2-10j(q+2)+5s = -(1+z)\frac{ds}{dz}-s(3+4q)
\end{align}

The equation of state (EoS) parameter for dark energy, denoted as $\omega_{DE}$, is expressed as:

\begin{equation}
\omega_{DE} = \frac{p_{DE}}{\rho_{DE}} = -1 + \frac{1+z}{3} \cdot \frac{2 \left[1 - \Omega_m(z)\right]}{H(z)^2} \cdot \frac{d H(z)^2}{dz}
\end{equation}

where $\Omega_m(z)$ represents the matter density parameter at a specific redshift, 
\begin{equation}
\Omega_m(z) = \frac{\Omega_{m0} (1+z)^3}{H(z)^2/H_0^2}
\end{equation}

\begin{table}[htbp]
\centering
\begin{tabular}{ll}
\toprule
\textbf{Dark Energy Type} & \textbf{EoS Parameter Range} \\
\midrule
Quintom Model & $\omega_{DE} > -1$, crossing $\omega_{DE} = -1$ line \\
Quintessence & $-1 < \omega_{DE} < -\frac{1}{3}$ \\
Phantom Model & $\omega_{DE} < -1$ \\
\bottomrule
\end{tabular}
\caption{Categorization of dark energy models using the EoS parameter.}
\label{tab:eos_range}
\end{table}

The stability of a cosmological model can be evaluated by examining the squared sound speed \cite{kim2008instability}:

\begin{equation}
v_s^2 = \frac{dp_{DE}}{d\rho_{DE}} = v^2(z) = \omega_{DE}(z) + \frac{(1+z)}{3} \frac{d \omega_{DE}(z)}{dz}
\end{equation}

For classical stability, $0 < v_s^2 < 1$.

The $\omega'-\omega$ phase plane, introduced by Caldwell and Linder \cite{PhysRevLett.95.141301}, differentiates dark energy models based on:

\begin{equation}
\omega'_{DE} = \frac{d}{d\chi} \omega_{DE}, \quad \text{where } \chi = \ln{a(t)}
\end{equation}

\begin{table}[htbp]
\centering
\begin{tabular}{lll}
\toprule
\textbf{Condition} & \textbf{Name} & \textbf{Remarks} \\
\midrule
$\omega'_{DE} > 0, \omega_{DE} < 0$ & Thawing Region & Less accelerated expansion \\
$\omega'_{DE} < 0, \omega_{DE} < 0$ & Freezing Region & More accelerated expansion \\
\bottomrule
\end{tabular}
\caption{Characteristics of $\omega'_{DE}-\omega_{DE}$ plot.}
\label{tab:omega_prime_omega}
\end{table}

The statefinder pair $(r,s^*)$, proposed by Sahni et al. \cite{sahni2003statefinder}, classifies dark energy independently of the model:

\begin{equation}
r = \frac{\dddot{a}}{aH^3}, \quad s^* = \frac{r - 1}{3(q - 1/2)}
\end{equation}

\begin{table}[htbp]
\centering
\begin{tabular}{lll}
\toprule
\textbf{Value} & \textbf{Type} & \textbf{Description} \\
\midrule
$(r < 1, s^* > 0)$ & Quintessence & Quintessence-like dark energy \\
$(r = 1, s^* = 0)$ & $\Lambda$CDM & $\Lambda$CDM dark energy \\
$(r > 1, s^* < 0)$ & Chaplygin Gas & Phantom-like dark energy \\
\bottomrule
\end{tabular}
\caption{Classification of dark energy models based on $(r,s^*)$ parameters.}
\label{tab:r_s_parameter}
\end{table}

Studies using these parameters to classify dark energy models can be found in \cite{chakraborty2012statefinder,panotopoulos2008statefinder,solanki2022statefinder, Alvarez_2022, Chang:2007jr,evans2005geometrical,alam2003exploring,feng2008statefinder}.

The $\text{OM}(z)$ diagnostic tool evaluates the normalised energy density relative to a flat matter-dominated universe:

\begin{equation}
\text{OM}(z) = \frac{E(z)^2 - 1}{(1 + z)^3 - 1}
\end{equation}

where $E(z) = H(z)/H_0$ denotes the normalized Hubble parameter. This diagnostic tool enables the exploration of the universe's expansion history in a model-independent manner:

\begin{itemize}
    \item $\text{OM}(z) > 0$: Matter-dominated universe
    \item $\text{OM}(z) < 0$: Dark energy influence
\end{itemize}

The redshift evolution of $\text{OM}(z)$ provides crucial diagnostic information regarding the cosmic transition from matter to dark energy dominance, facilitating constraints on dark energy's equation of state and illuminating the nature of universal expansion.

The temporal duration of cosmic evolution, denoted as $t_{\text{age}}$, can be mathematically expressed through an integration of the reciprocal Hubble parameter $H(z)$ across the redshift domain. This calculation spans from the present epoch ($z = 0$) to primordial times ($z \to \infty$), taking the mathematical form:

\begin{equation}\label{universe_age_calculation}
t_{\text{age}} = \int_{0}^{\infty} \frac{dz}{(1+z)H(z)}
\end{equation}

Within this formulation, $H(z)$ represents the Hubble expansion rate, while the $(1+z)$ term accounts for relativistic time dilation effects arising from cosmic expansion. The evaluation of this integral depends fundamentally on the adopted cosmological framework, incorporating contributions from various cosmic components including matter, radiation, and dark energy. This mathematical approach enables precise age determinations across diverse cosmological scenarios.

\section{An Overview of Einstein-Aether Gravity} \label{Sect: EA Model}
Einstein-aether gravity is a well-established vector-tensor theory that extends the framework of general relativity by incorporating a dynamical, time-like unit vector field, referred to as the ``aether,'' which locally selects a favored reference frame at each point in spacetime. The initial development of Einstein-Aether gravity was further advanced by Gasperini through a series of papers, where the theory was popularized as a compelling alternative to existing models \cite{gasperini1987singularity}. The formal vector-tensor framework, initially proposed by Jacobson and Mattingly \cite{jacobson2001gravity,jacobson2004einstein,jacobson2008einstein}, was later generalized by Zlosnik, Ferreira, and Starkman \cite{zlosnik2007modifying,zlosnik2008growth}. This concept revisits the classical notion of aether—a medium once hypothesized to permeate the universe, serving as a reference frame for absolute motion. However, unlike the mechanical aether of pre-relativistic physics, the aether in this theory is not a static or material entity. Instead, it is a dynamical vector field that coexists with the spacetime metric, determining local physical conditions while maintaining general covariance. A primary motivation for Einstein-Aether gravity is to introduce a preferred frame while maintaining the principles of general covariance and the integrity of Einstein's field equations. In general relativity, general covariance pertains to the decreasing divergence of the energy-momentum tensor, ensuring consistency with the contracted Bianchi identity. The aether field in Einstein-aether gravity, by being dynamic rather than fixed, preserves this covariance by allowing the preferred frame to evolve with the spacetime geometry. This theory serves as a phenomenological probe of possible Lorentz violations, which are anticipated in certain approaches to quantum gravity. Moreover, Einstein-aether gravity provides a compelling framework for exploring modifications to gravitational dynamics, particularly in the context of explaining phenomena typically attributed to dark matter \cite{zlosnik2007modifying}. By introducing the aether as a dynamical vector field, the theory can potentially explain galactic rotation curves and other cosmological observations without resorting to additional dark matter components. This methodology is consistent with the wider category of modified gravity theories, which adjust the gravitational interaction at large scales to address cosmological anomalies. Consequently, Einstein-aether gravity is of significant interest for its ability to address fundamental questions in cosmology and quantum gravity, offering an alternative perspective on the nature of spacetime and gravitational interactions. 
\par The action of the generalised Einstein-Aether gravity can be written as,
\begin{equation}\label{EA_action}
    S = \int d^4x \sqrt{-g}\left[\dfrac{R}{16\pi G}+\mathcal{L}_M + \mathcal{L}_A \right]
\end{equation}
Within this mathematical framework, we define $R$ as the Ricci scalar curvature and $G$ as Newton's gravitational constant. The total system dynamics are governed by two distinct Lagrangian densities: $\mathcal{L}_M$ for the matter field contributions and $\mathcal{L}_A$ describing the aether field behavior. The aether field's Lagrangian density incorporates quadratic terms in both the field and its derivatives, taking the following form \cite{zlosnik2007modifying}:

\begin{gather}\label{L_A value}
    \mathcal{L}_A = \frac{M^2}{16\pi G}F(K) + \frac{1}{16\pi G}\lambda(A^{\alpha}A_{\alpha} + 1) \\
    K = M^{-2} K^{\alpha \beta}_{\gamma \sigma} \nabla_{\alpha} A^{\gamma} \nabla_{\beta} A^{\sigma} \\
    K_{\gamma \sigma}^{\alpha \beta} = c_1 g^{\alpha \beta} g_{\gamma \sigma} + c_2 \delta_{\gamma}^{\alpha} \delta_{\sigma}^{\beta} + c_3 \delta_{\sigma}^{\alpha} \delta_{\gamma}^{\beta}
\end{gather}
The parameters $c_i$ represent dimensionless coupling constants, while $M$ serves as a mass-dimensional coupling parameter. The theory incorporates an undetermined function $F(K)$, which remains a free parameter of the model. To maintain the unit norm condition for the time-like vector field, we introduce a Lagrange multiplier term $\lambda$. It's worth noting that some theoretical treatments incorporate an additional coupling term $c_4 A^\alpha A^\beta g^{\gamma\sigma}$ in the definition of $K_{\alpha\beta}^{\gamma\sigma}$ \cite{jacobson2004einstein}. In deriving the field equations, we treat both the inverse metric tensor $g^{\alpha\beta}$ and the contravariant vector field $A^\beta$ as independent dynamical variables. The complete set of equations of motion emerges from varying the action in~\eqref{EA_action} with respect to these fundamental variables, yielding both the gravitational field equations and the vector field evolution equation.

\begin{gather}\label{EA_fieldequations}
    G_{\alpha\beta} = T_{\alpha\beta}^A + 8\pi G T_{\alpha\beta}^M \\
    \nabla_{\alpha} (F_K J^{\alpha}_{\ \beta}) = 2\lambda A_{\beta}
\end{gather}
In this context, \( T_{\alpha\beta}^A \) denotes the stress-energy tensor associated with the vector field aether, while \( T_{\alpha\beta}^M \) represents the stress-energy tensor associated with the matter fields and,
\begin{gather*}
    F_K = \frac{dF}{dK} \\
    J^{\alpha}_{\sigma} = 2K^{\alpha \beta}_{\sigma \gamma} \nabla_{\beta} A^{\gamma}
\end{gather*}
Using equations \eqref{L_A value} we can obtain,
\begin{equation}\label{EA_stress_energy_tensor}
    T_{\alpha \beta}^A = \frac{1}{2} \nabla_{\sigma} \left[ F_K \left( J^{\sigma}_{(\alpha} A_{\beta)} - J^{\sigma}_{(\alpha} A_{\beta)} - J_{(\alpha \beta)} A^{\sigma} \right) \right] - F_K Y_{(\alpha \beta)} + \frac{1}{2} g_{\alpha \beta} M^2 F + \lambda A_{\alpha} A_{\beta}
\end{equation}
where the subscript \((\alpha\beta)\) indicates the symmetry concerning the indices, and
\begin{equation}
   Y_{\alpha \beta} = -c_1 \left[ (\nabla_{\nu} A_{\alpha})(\nabla^{\nu} A_{\beta}) - (\nabla_{\alpha} A_{\nu})(\nabla_{\beta} A^{\nu}) \right]
\end{equation}
Moreover, the condition that \( A \) is a unit vector field that is time-like enforces the constraint \( A^{\alpha}A_{\alpha} = -1 \). We will now examine the scenario of a homogeneous and isotropic universe, which is supported by the WMAP observations, represented by the Friedmann-Robertson-Walker (FRW) metric:
\begin{equation}\label{FRW_metric}
    ds^2 = -dt^2 + a(t)^2 \left( \frac{dr^2}{1 - kr^2} + r^2 d\Omega^2 \right),
\end{equation}

The scale factor of the universe is represented by $a(t)$, while the spatial curvature parameter is denoted by $k$, and the spherical component of the metric is expressed through $d\Omega^2$. Given the fundamental requirements of cosmic homogeneity and isotropy at large scales, the vector field must maintain consistency with these symmetries. This symmetry constraint, combined with the normalization condition $A^\alpha A_\alpha = -1$, naturally leads to a simplified vector field configuration \cite{meng2012specific}:

\begin{equation}\label{A_value}
    A^\alpha = (1, 0, 0, 0)
\end{equation}

For the material content of the universe, we adopt the perfect fluid description, characterized by an energy-momentum tensor with the structure:

\begin{equation}\label{stress_energy_matter}
T^M_{\alpha\beta} = \rho U_\alpha U_\beta + p (U_\alpha U_\beta + g_{\alpha\beta})
\end{equation}

In this expression, $U^\alpha$ represents the four-velocity field of the cosmic fluid. The kinetic term $K$ can be evaluated by combining the geometric properties encoded in \eqref{FRW_metric} with the vector field configuration specified by \eqref{A_value}.

\begin{equation}\label{K_value}
K = M^{-2} (c_1 g^{\alpha\beta} g_{\gamma\sigma} + c_2 \delta^\alpha_\gamma \delta^\beta_\sigma + c_3 \delta^\alpha_\sigma \delta^\beta_\gamma) = \frac{3\beta H^2}{M^2},
\end{equation}
where the coefficient is defined as \( \beta = c_1 + 3c_2 + c_3 \), and H, the Hubble parameter is expressed as \( H \equiv \dot{a}/a \). Furthermore, It is noted that the vector field's stress-energy tensor assumes the characteristics of a perfect fluid, whose energy density is represented by~\cite{zlosnik2008growth}
\begin{equation}\label{EA_energydensity}
    \rho_{EA} = 3\beta H^2 \left[F_K-\dfrac{F}{2K}\right]
\end{equation}
The pressure is given by
\begin{equation}\label{EA_pressure}
    p_{EA}= -3\beta H^2 \left[F_K-\dfrac{F}{2K}\right]-\beta\left(\dot{H}F_K+H\dot{F}_K\right)
\end{equation}
Here, dots indicate the time derivative. We can confirm that the contributions from the vector field component fulfill the energy conservation relation in cosmology,

\begin{equation}
\dot{\rho}_{EA} + 3H(\rho_{EA} + p_{EA}) = 0.
\end{equation}
This conservation relation holds for any form of $F(K)$ \cite{meng2012specific}. The modified Friedmann equations are given using \eqref{EA_fieldequations} as,
\begin{equation}\label{EA_Friedmannequations}
H^2 + \frac{k}{a^2} = \frac{8\pi G}{3} \rho + \frac{1}{3} \rho_{EA},
\end{equation}

\begin{equation}
-2\dot{H} + \frac{2k}{a^2} = 8\pi G (\rho + p) + (\rho_{EA} + p_{EA}),
\end{equation}
The equation of state (EoS) parameter corresponding to the Einstein-Aether contribution is represented as

\begin{equation}\label{EA_EoS_parameter}
w_{EA} = \frac{p_{EA}}{\rho_{EA}} = -1 - \frac{\dot{H}F_K + \dot{F}_K H}{3H^2 \left( F_K - \frac{F}{2K} \right)}.
\end{equation}

Given that the density remains positive at all times, the condition \( \rho_{EA} > 0 \) implies that \( F_K > \frac{F}{2K} \), assuming \( \beta > 0 \). The effective density and pressure derived from Einstein-Aether gravity can yield dark energy for the condition~\cite{debnath2014reconstructions}

\begin{equation}
\rho_{EA} + 3p_{EA} < 0
\end{equation}

is satisfied. This leads to the inequality 

\begin{equation}
2H^2 \left( F_K - \frac{F}{2K} \right) > -\left( \dot{H}F_K + \dot{F}_KH \right).
\end{equation}
The temporal evolution of the universe can be characterized through a power-law formulation of the scale factor \cite{nojiri2007introduction}:

\begin{equation*}
    a(t) = a_0 t^\delta
\end{equation*}

where $a_0$ denotes the present-day value of the scale factor, and the power-law index satisfies $\delta > 0$. This parametrization enables us to derive explicit expressions for several key cosmological quantities: the Hubble expansion rate $H$, its temporal derivative $\dot{H}$, and the term $K$.

\begin{equation}\label{H,K,dotH value}
H = \frac{\delta}{t}, \quad \dot{H} = -\frac{\delta}{t^2}, \quad K = \frac{3\beta \delta^2}{M^2 t^2}.
\end{equation}
The connection between the scale factor \(a(t)\) and the redshift parameter \(z\) is given by:
\begin{equation}\label{redshift_a(t)}
    1+z = \frac{1}{a(t)}
\end{equation}

\section{Barrow Agegraphic Dark Energy} \label{Sect: BADE Model}
When extended to a cosmological context, the holographic principle posits that the entropy connected to the horizon of the universe (the maximum observable scale) is proportional to its surface area and the entropy resembles the Bekenstein-Hawking entropy formula for black holes~\cite{bousso2002holographic, hooft1993dimensional, Susskind1994TheWA, fischler1998holography}. For further discussions on using the holographic principle to describe dark energy, the literature in \cite{Cohen_1999, Enqvist_2005, Gong_2004, Pavo_n_2006, ZHANG_2005} offers more insights.

\par Motivated by the complex surface structure of the COVID-19 virus, Barrow recently proposed a novel entropy formulation for black holes, accounting for quantum-gravitational effects that can imply complex, fractal-like features on the black hole event horizon. This modification yields a finite volume but an infinitely large (or significantly enlarged) area, thus altering the conventional entropy-area relationship. The Barrow entropy is given by \cite{Barrow_2020}:
\begin{equation}\label{Barrowentropy}
    S_B = \left(\frac{A}{A_0}\right)^{\left(1 + \frac{\Delta}{2}\right)},
\end{equation}
The fundamental parameters in this formulation include the horizon surface area $A$, the fundamental Planck area $A_0$, and the quantum-gravitational deformation parameter $\Delta$ (known as the Barrow exponent). The parameter $\Delta$ serves as an indicator of horizon complexity, where $\Delta = 0$ recovers the classical Bekenstein-Hawking entropy formulation, while $\Delta = 1$ indicates maximum complexity in the horizon structure.nThe extension of this entropy formalism to dark energy physics builds upon the fundamental holographic principle, which establishes an upper bound on the dark energy density through the relation $\rho_{DE}L^4 \leq S$ \cite{Wang_2017}. This constraint, when combined with the modified entropy expression in Eq.~\eqref{Barrowentropy}, leads to a novel formulation of dark energy density \cite{Saridakis_2020}:

\begin{equation}\label{rho_BHDE}
    \rho_{BHDE} = C L^{\Delta - 2}
\end{equation}

Here, $C$ represents a dimensional parameter with units $[L]^{-\Delta - 2}$, and $L$ characterizes the relevant horizon scale. The implementation of Barrow entropy within the holographic framework, utilizing the future event horizon as an infrared (IR) cut-off, was first proposed as the Barrow Holographic Dark Energy model \cite{Saridakis_2020}. This framework was subsequently expanded by the introduction of an alternative formulation employing the Hubble horizon as the IR cut-off, accompanied by detailed analysis of its cosmological implications \cite{srivastava2021barrow}. In the limiting case where $\Delta$ vanishes, Eq.~\eqref{rho_BHDE} reduces to the standard holographic dark energy expression $\rho_{DE} = 3c^2M_P^2L^{-2}$, with the identification $C = 3c^2M_P^2$, where $M_P$ denotes the Planck mass and $c$ is a dimensionless parameter. The non-zero values of $\Delta$ introduce modifications to conventional holographic dark energy behavior, resulting in distinct evolutionary patterns in cosmological dynamics.

\par In the framework of holographic dark energy theories, the characteristic length scale $L$ emerges as a fundamental parameter that delineates the boundary where quantum gravitational phenomena become relevant. This characteristic length introduces constraints on the dark energy density, thereby influencing cosmic evolutionary dynamics. The theoretical foundation of holographic principles in gravitational physics received substantial reinforcement through the groundbreaking AdS/CFT correspondence \cite{Maldacena_1999}. This fundamental duality establishes a mathematical connection linking gravitational dynamics in anti-de Sitter (AdS) spacetime to conformal field theories operating on the spacetime boundary. Such a correspondence illuminates the profound role of holographic length scales in connecting gravitational phenomena in the bulk with quantum field theories on boundaries. The relationship between holographic principles and gravitational dynamics gained additional perspective through the development of ``emergent gravity'' theory \cite{Verlinde_2011}. This theoretical framework proposes that gravitational interactions may emerge from more fundamental entropic forces, which are intrinsically connected to information dynamics on holographic surfaces. Within this paradigm, the holographic length scale assumes a fundamental role in describing how gravitational phenomena emerge from underlying information-theoretic principles.

\par Quantum properties of spacetime find representation in the agegraphic dark energy paradigm, which proposes a correlation between dark energy density and universal age. This theoretical construct offers an alternative that resolves conceptual difficulties associated with future event horizons present in holographic frameworks. Drawing from foundational works \cite{cai2007dark, wei2008new}, the energy density in this model takes the form:

\begin{equation}\label{ADE_rho}
   \rho_{ADE} = \frac{3n^2 M_P^2}{T^2},
\end{equation}

where:
- $T$ represents the temporal age of the universe
- $M_P$ denotes the reduced Planck mass
- $n$ serves as a dimensionless parameter

The selection of universal age or conformal time as infrared (IR) cutoff parameters carries fundamental significance for several theoretical considerations:

\begin{itemize}
    \item The selection of a time-related scale connects dark energy to the large-scale quantum fluctuations of spacetime \cite{huang2004holographic}.
    \item It avoids causality issues associated with the future event horizon by focusing on observable quantities like the universe's age.
    \item The model naturally explains late-time cosmic acceleration, as dark energy becomes relevant only in the universe's later stages \cite{wei2008new}.
    \item Conformal time as an IR cutoff better captures the causal structure of the universe, aligning dark energy with the full spacetime evolution \cite{ZHANG_2005}.
\end{itemize}
A novel approach to dark energy modeling was introduced by Sharma, Varshney, and Dubey \cite{Sharma_2021}, who formulated two distinct frameworks: the Barrow Agegraphic Dark Energy and New Barrow Agegraphic Dark Energy models. These frameworks distinguish themselves through their respective infrared (IR) cutoff selections, with BADE utilizing the universe's age \(T\) and NBADE employing the conformal time \(\eta\). While the BADE model applies Barrow entropy corrections to the agegraphic dark energy framework by using the cosmic age $T$ as the infrared (IR) cutoff, the NBADE model replaces this with the conformal time $\eta$. This change is not arbitrary — it is motivated by the causal structure of spacetime, where conformal time more naturally captures the horizon evolution and the propagation of quantum fields in a curved background. In particular, conformal time plays a crucial role in early-universe physics, making NBADE better suited for probing dark energy behavior across a wider redshift range, including the matter-dominated and early-accelerating epochs. From an entropic perspective, choosing $\eta$ as the IR cutoff allows Barrow corrections to be incorporated into a more causally consistent and scale-sensitive formulation of holographic dark energy. Conformal time plays a pivotal role in inflationary dynamics, cosmic microwave background (CMB) perturbation theory, and horizon-scale quantum processes. In contrast to the cosmic age $T$, which merely counts elapsed time, conformal time measures the true causal span accessible to massless particles. Furthermore, in the early universe, the integral $\eta = \int dt/a(t)$ naturally emphasizes high-redshift behavior, allowing NBADE to more accurately model pre-accelerated epochs and early DE evolution. 

The combined framework of conformal time as an IR cutoff and Barrow entropy corrections provides a more coherent theoretical picture when viewed through the lens of emergent gravity theories \cite{Verlinde_2011}. In these approaches, gravity emerges as an entropic force related to information holographically encoded on surfaces. The conformal time naturally connects to the lightcone structure of spacetime, which is fundamental to defining causal relationships and information bounds in quantum gravitational frameworks.
Moreover, recent developments in quantum gravity approaches such as causal set theory \cite{sorkin2005causal} and causal dynamical triangulations \cite{ambjorn2005semiclassical} have emphasized the importance of causal structure in quantum spacetime, which aligns more naturally with the conformal time parameter than with proper time. The conformal time captures the maximum potential causal influence throughout cosmic history, making it a more fundamental quantity from an information-theoretic perspective.
When these quantum gravitational considerations are combined with Einstein-Aether theory's introduction of a preferred timelike direction, the NBADE framework offers a unique opportunity to probe potential signatures of quantum gravity in cosmological observations. The reconstruction of Einstein-Aether gravity from NBADE thus represents not merely an exercise in fitting cosmological data but a deeper exploration of how fundamental quantum gravitational principles might manifest in the large-scale structure and evolution of the universe. The cosmic age and conformal time are mathematically expressed as:

\begin{equation}\label{universe_age}
   T = \int_0^t dt = \int_0^a \dfrac{da}{Ha}
\end{equation}

\begin{equation}\label{conformal_time}
   \eta = \int_0^a \dfrac{da}{a^2H}
\end{equation}

where the Hubble parameter is denoted by $H$ and the cosmic scale factor by $a$. Through application of Barrow entropic principles, the energy density expressions for these models emerge as:

\begin{equation}\label{rho_BADE}
   \rho_{\text{BADE}} = CT^{\Delta-2}
\end{equation}

\begin{equation}\label{rho_NBADE}
   \rho_{\text{NBADE}} = C\eta^{\Delta-2}
\end{equation}

In these expressions, $T$ represents the universal age parameter, while $\eta$ denotes the conformal time measure. To provide a clearer theoretical motivation for studying both BADE and NBADE within the Einstein-Aether framework, a comparative analysis of their key characteristics is given in Tab.~\ref{tab:bade_nbade_comparison}:

\begin{table}[ht]
\centering
\caption{Comparative analysis of BADE and NBADE characteristics}
\label{tab:bade_nbade_comparison}
\begin{tabular}{|p{4.5cm}|p{5cm}|p{5cm}|}
\hline
\textbf{Aspect} & \textbf{BADE (Age Cutoff)} & \textbf{NBADE (Conformal Time Cutoff)} \\
\hline
\textbf{Causal Structure} & Limited to proper time measurements & Captures full light-cone causal structure \\
\hline
\textbf{Matter Era Compatibility} & Challenges in reproducing matter-dominated epoch & Successfully accommodates matter-dominated phase \\
\hline
\textbf{Theoretical Consistency} & May require additional mechanisms for early universe & Naturally connects early and late-time dynamics \\
\hline
\textbf{Quantum Gravity Alignment} & Limited connection to quantum causal structure & Better aligned with quantum spacetime frameworks \\
\hline
\textbf{Einstein-Aether Compatibility} & Aligns with preferred time direction & Provides richer connection to spacetime causal structure \\
\hline
\end{tabular}
\end{table}

This comparison reveals that NBADE is not merely an ad hoc variant of BADE but represents a theoretically motivated refinement that addresses specific limitations of the original formulation. The conformal time cutoff in NBADE offers a more comprehensive framework for understanding dark energy dynamics across all cosmic epochs, while maintaining the quantum-gravitational insights provided by Barrow entropy. Like ADE\cite{Zhang:2007ps}, depends on the cosmic age $T$, which varies significantly across different epochs, making the model strongly epoch-dependent and less suitable as a universal framework. In the early universe, it behaves like a cosmological constant but does not dominate; during the matter-dominated era, its scaling behavior creates inconsistencies, preventing it from naturally overtaking matter. In contrast, NBADE functions effectively as a single-parameter model, with all physical quantities determined by the parameter $n$, offering analytical simplicity and greater predictive power\cite{wu2007observational}. Furthermore, when $n$ is of order unity, NBADE inherently addresses the cosmological coincidence problem, providing a more consistent and observationally viable alternative\cite{Wei:2007xu}. This theoretical advantage makes the reconstruction of Einstein-Aether gravity from NBADE particularly compelling, as it combines three conceptually aligned elements: modified gravity with a preferred time direction (Einstein-Aether), quantum-gravitational deformations of entropy (Barrow entropy), and a cosmologically complete causal structure (conformal time cutoff). The comparison between reconstructions from BADE and NBADE within the same Einstein-Aether framework allows us to evaluate which IR cutoff better captures the underlying physics of dark energy when both observational constraints and theoretical consistency are considered. This approach provides a robust methodology for discriminating between competing dark energy models based on both empirical evidence and theoretical foundations. Our investigation proceeds under the assumption of non-interacting dark matter and dark energy components.

\subsection{\bf Reconstruction from BADE} \label{Sect: Reconstruction BADE}
\begin{figure}[htbp]
    \centering
    \begin{subfigure}[b]{0.45\textwidth}
        \includegraphics[width=\textwidth]{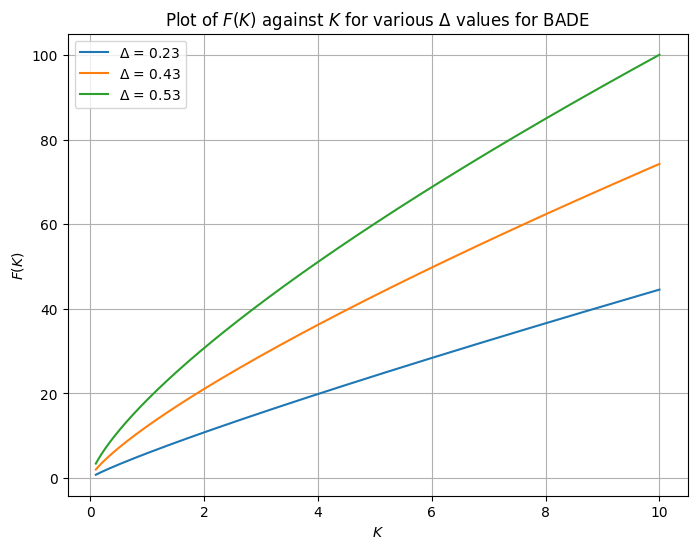}
        \caption{}
        \label{fig:f(K)_K_BADE}
    \end{subfigure}
    \begin{subfigure}[b]{0.45\textwidth}
        \includegraphics[width=\textwidth]{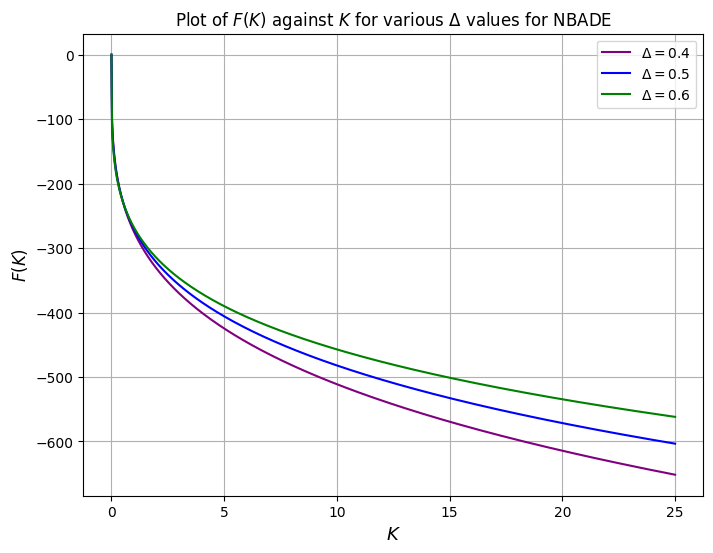}
        \caption{}
        \label{fig:f(K)_K_NBADE}
    \end{subfigure}
    \caption{The plot of \( F(K) \) versus \( K \) is presented for the reconstruction of Einstein-Aether gravity, as shown in Fig. (a) for the BADE model and Fig. (b) for the NBADE model.}
    \label{fig:f(K)_K_BADE_NBADE}
\end{figure}

Here, we will derive the unknown function $F(K)$ from BADE. Now, using the relations in \eqref{H,K,dotH value} and \eqref{universe_age} we obtained,
\begin{equation}\label{T_expression}
    T = \left(\dfrac{3\beta\delta^2}{M^2K}\right)^{\dfrac{1}{2}}
\end{equation}
The energy density of BADE can be re-written as:
\begin{equation}\label{rho_BADE_Tvalue}
    \rho_{\text{BADE}}=C\left(\dfrac{3\beta\delta^2}{M^2K}\right)^{\dfrac{\Delta-2}{2}}
\end{equation}
To derive Einstein-Aether gravity from Barrow Agegraphic Dark Energy, we establish an analogy by equating the dark energy densities given in \eqref{EA_energydensity} and \eqref{rho_BADE_Tvalue}:
\begin{equation}\label{BADE_reconstruction_equation}
    3\beta H^2 \left[F_K-\dfrac{F}{2K}\right] = C\left(\dfrac{3\beta\delta^2}{M^2K}\right)^{\dfrac{\Delta-2}{2}}
\end{equation}
The solution of this equation gives,
\begin{equation}\label{f(K)_BADE}
    \boxed{F(K) = -\dfrac{2 \cdot 3^{(\frac{\Delta}{2}-1)} C \left(\dfrac{\beta \delta^2}{K M^2}\right)^{\frac{1}{2}(\Delta-2)}}{M^2 (\Delta-1)} + \sqrt{K} C_1}
\end{equation}
Here, $C_1$ is the constant of integration chosen arbitrarily. This gives the reconstructed expression of the unknown function $F(K)$. The above expression~\eqref{f(K)_BADE} is analytical for $M \neq 0$, $K \neq 0$ and $\Delta \neq 1$. The condition that $\Delta$ should not be equal to $1$ for the reconstructed $F(K)$ to be analytical imposes a constraint on the Barrow exponent $\Delta$. In Fig.~\ref{fig:f(K)_K_BADE}, we present the reconstructed function \( F(K) \) plotted against \( K \) for three distinct values of \( \Delta \). The results demonstrate that \( F(K) \) is an increasing function of \( K \) for all values of \( \Delta \).
\subsection{\bf Reconstruction from NBADE} \label{Sect: Reconstruction NBADE}
Now, we will reconstruct the Einstein-Aether gravity from NBADE. From the relations given in \eqref{conformal_time}, \eqref{H,K,dotH value} and taking the above defined scale factor relation $a(t)=a_0t^{\delta}$, the conformal time $\eta$ can be written as,
\begin{equation}\label{eta_expression}
    \eta = \int_0^t \frac{dt}{a} = \frac{t^{(1-\delta)}}{a_0(1-\delta)}
\end{equation}
Then, the energy density of NBADE is given by,
\begin{equation}\label{rho_NBADE_etavalue}
    \rho_{\text{NBADE}}=C \left[ a_0 (1 - \delta) \right]^{2 - \Delta} \left( \frac{3 \beta \delta^2}{K M^2} \right)^{\frac{(1 - \delta)(\Delta - 2)}{2}}
\end{equation}
Now, to reconstruct the unknown function $F(K)$ from NBADE we have to equate the corresponding energy densities given in \eqref{EA_energydensity} and \eqref{rho_NBADE_etavalue},
\begin{equation}\label{NBADE_reconstruction_equation}
    3\beta H^2 \left[F_K-\dfrac{F}{2K}\right] = C \left[a_0 (1 - \delta) \right]^{2 - \Delta} \left( \frac{3 \beta \delta^2}{K M^2} \right)^{\frac{(1 - \delta)(\Delta - 2)}{2}}
\end{equation}
Integrating the equation \eqref{NBADE_reconstruction_equation} we obtain,
\begin{equation}\label{f(K)_NBADE}
    \boxed{F(K) = \dfrac{{2 \cdot 3^{(0.5 - 0.5\delta)(\Delta - 2)}  C  a_0^2  \left(\dfrac{\beta \delta^2}{K M^2}\right)^{(0.5 - 0.5\delta)(\Delta - 2)} (\delta - 1)^2}}{{M^2 (-a_0\delta + a_0)^\Delta (-\Delta + \delta(\Delta - 2) + 1)}} + C_2 \sqrt{K}}
\end{equation}
Here, $C_2$ is the integration constant. The equation \eqref{f(K)_NBADE} gives the reconstructed expression of $F(K)$ from NBADE. The above function $F(K)$ is analytic except $M=0$, $K=0$, $\delta=1$ and $\Delta = \dfrac{1-2\delta}{\delta-1}$. Thus reconstruction of Einstein-Aether gravity from NBADE is feasible as finite number of singularities are there. We have shown the plot of the reconstructed $F(K)$ against $K$ in the fig.~\ref{fig:f(K)_K_NBADE} for several values of $\Delta$. As it is seen from the plot, the function $F(K)$ is decreasing with the increase of $K$ when it is reconstructed from New Barrow Agegraphic Dark Energy.

\section{Methodology} \label{Sect: Methodology}
In the part that follows, we will provide the observational datasets and methodology to restrict the free parameters of reconstructed Einstein-Aether gravity from BADE and NBADE models. We implemented the MCMC algorithm in Python using the \texttt{emcee} package \cite{foreman2013emcee}. The likelihood function was defined by assuming Gaussian errors in the observational data. Uniform priors were assigned to the model parameters within physically motivated bounds. 

\begin{figure}[htbp]
    \centering
    \begin{subfigure}[b]{0.55\textwidth} % Adjusted width
        \includegraphics[width=1.3\textwidth, height=1.0\textwidth]{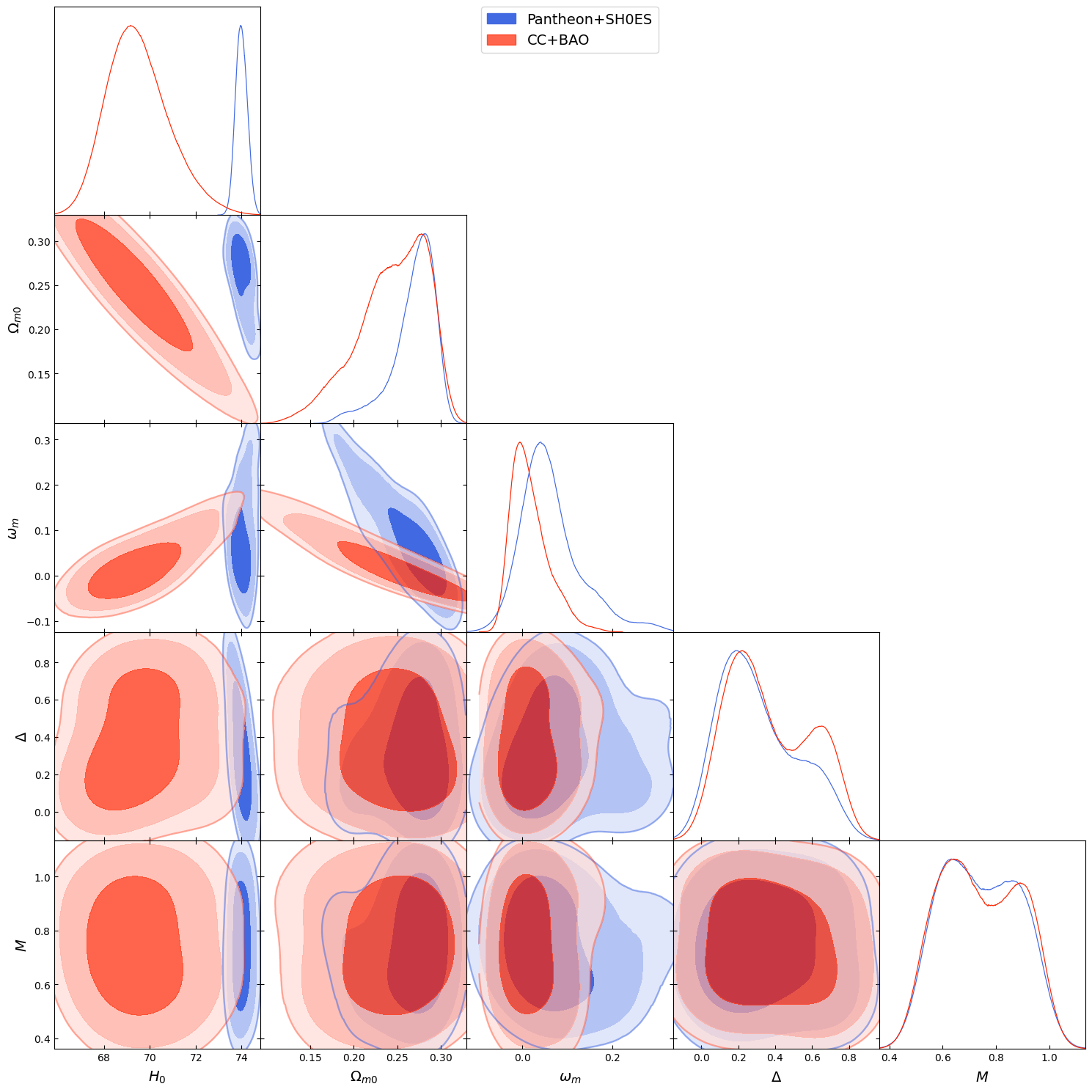} % Adjusted height
        \caption{}
        \label{fig:getdist_BADE}
    \end{subfigure}
    \hfill
    \begin{subfigure}[b]{0.55\textwidth} % Adjusted width
        \includegraphics[width=1.3\textwidth, height=1.0\textwidth]{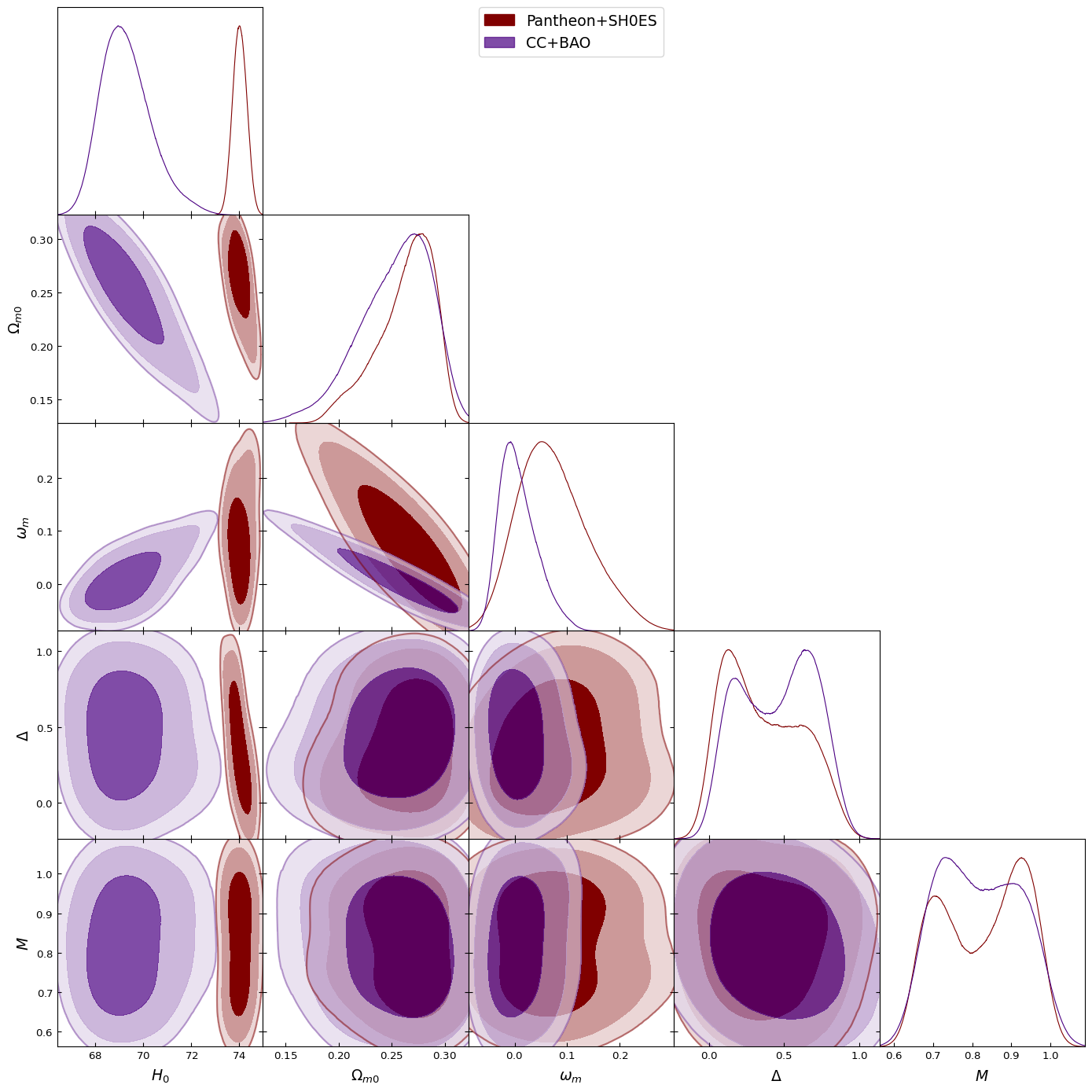} % Adjusted height
        \caption{}
        \label{fig:getdist_NBADE}
    \end{subfigure}
    \caption{Posterior distribution of reconstructed Einstein-Aether gravity from fig: (a) BADE Model and fig: (b) NBADE Model at $1\sigma$ and $2\sigma$ confidence level.}
    \label{fig:getdist_BADE_NBADE}
\end{figure}

Markov Chain Monte Carlo (MCMC) simulations were conducted with $10000$ iterations, preceded by a $100$-step burn-in period to establish convergence. A swarm of $100$ walkers was utilized to explore the parameter space efficiently. The Gelman-Rubin diagnostic was monitored to verify convergence, ensuring that $R$ remained below $1.1$ for all parameters. The optimal parameter values derived from our analysis are summarized in Table~\ref{tab: best_fit_params_values}. The posterior distributions at 1$\sigma$ and 2$\sigma$ confidence levels are illustrated in Figure~\ref{fig:getdist_BADE_NBADE}. For the BADE model, the constant parameters were fixed at $a_0=0.88$, $\beta=1$, $\delta=5$, $C_1 = 10$, and $C=100$. In contrast, the NBADE model employed $a_0 = 1.76$, $\beta=1$, $\delta=0.65$, $C_2=100$, and $C=100$.

\subsection{Cosmic Chronometers(CC) and Baryon Acoustic Oscillation(BAO) Dataset:}
Our analysis employs a comprehensive observational dataset combining 26 Baryon Acoustic Oscillation (BAO) measurements with 31 Cosmic Chronometer (CC) observations, yielding 57 distinct cosmological constraints. The CC methodology provides a direct measurement technique for the Hubble parameter $H(z)$ at varying redshifts. This approach evaluates differential age measurements between passive galaxies characterized by evolved stellar populations and minimal ongoing star formation. A significant advantage of CC observations lies in their cosmological model independence, enabling unbiased investigation of cosmic expansion dynamics. Our utilized CC dataset encompasses 31 $H(z)$ measurements distributed across redshifts from $z \approx 0.07$ to $z \approx 1.97$. These measurements derive from multiple investigations \cite{stern2010cosmic, moresco2012new, moresco2012improved, zhang2014four, moresco2015raising, moresco20166}, employing differential age analysis of early-type galactic systems.

The BAO phenomenon manifests as a primordial acoustic signature imprinted in galactic distribution patterns. This characteristic scale serves as a cosmic standard ruler, facilitating precise determinations of both angular diameter distances and Hubble parameters across cosmic time. Our investigation incorporates 26 BAO measurements, encompassing both isotropic and anisotropic observations from multiple galaxy surveys including SDSS \cite{ross2015clustering, alam2017clustering, gil2020completed, raichoor2021completed, hou2021completed, des2020completed}, DES (Dark Energy Survey) \cite{abbott2022dark}, DECaLS (Dark Energy Camera Legacy Survey) \cite{sridhar2020clustering}, and 6dFGS BAO \cite{beutler20116df}. These observations span a redshift interval of $0.106<z<2.36$. 

For statistical analysis of the combined CC+BAO dataset, we implement a chi-square likelihood formalism:

$$\chi^2_{\text{CC+BAO}} = -\frac{1}{2}\sum_i \left(\frac{H\text{obs}(z_i) - H_\text{model}(z_i)}{\sigma_i}\right)^2$$
where \( H_\text{obs}(z_i) \) represents the observed Hubble parameter at redshift \( z_i \), \( H_\text{model}(z_i) \) denotes the predicted value from the model, and \( \sigma_i \) indicates the associated uncertainty.

\subsection{Pantheon+SH0ES Supernova Dataset:}
The comprehensive analysis of cosmic expansion has been revolutionized through the combination of two significant datasets: the expanded Pantheon+ Type Ia supernovae (SNe Ia) catalog and the high-precision Hubble constant measurements from the SH0ES (Supernovae and \(H_0\) for the Equation of State) collaboration. The enhanced Pantheon+ compilation encompasses observations of 1701 SNe Ia distributed across an extensive redshift interval (\(0.01 \leq z \leq 2.3\)). These observations were accumulated using an array of observational platforms, combining both terrestrial facilities and space-based instruments, notably the Hubble Space Telescope (HST). A significant advancement in the Pantheon+ dataset lies in its refined calibration procedures, which effectively minimize systematic uncertainties and strengthen the robustness of cosmological constraints~\cite{Scolnic_2018, Scolnic_2022}.

The SH0ES methodology employs Cepheid variables as distance indicators to galaxies containing SNe Ia, establishing a fundamental calibration for cosmic distance measurements. This approach has yielded a precise determination of the Hubble constant, with \(H_0 = 73.04 \pm 1.04 \ \text{km s}^{-1} \text{Mpc}^{-1}\). This value exhibits a notable discrepancy when compared with predictions derived from Cosmic Microwave Background (CMB) observations, potentially indicating physics beyond the standard \(\Lambda\)CDM cosmological framework~\cite{Riess_2022, Riess2021}.

The synergy between Pantheon+ and SH0ES observations creates an unprecedented framework for examining the universe's expansion dynamics, particularly focusing on the phenomenon of cosmic acceleration in recent epochs. This combined dataset enables robust constraints on fundamental cosmological parameters, including the redshift-dependent Hubble parameter \(H(z)\), the dark energy equation of state \(\omega_{DE}\), and various other diagnostics characterizing the nature of dark energy. A fundamental relationship in this analysis connects the distance modulus \(\mu(z)\) to the luminosity distance \(d_L(z)\).

\[
\mu(z) = 5 \log_{10} \left( d_L(z) \right) + 25
\]
The luminosity distance, expressed as $d_L(z)=(1+z)D_M$, emerges from the numerical solution of coupled differential equations involving the redshift $z$ and $d_L(z)$. In the context of a spatially flat universe, the comoving distance $D_M$ takes the mathematical form:

\[D_M = \frac{c}{H_0}\int_0^z \frac{dz'}{E(z')}\]

In this expression, $c$ represents the universal constant of light propagation in vacuum, while $H_0$ quantifies the present-day rate of cosmic expansion.

For statistical analysis of the Pantheon+SH0ES observations, we implement a log-likelihood function structured as:

\[
\mathcal{L}_{\text{SN}} = -\frac{1}{2} \left( \boldsymbol{\mu}_{\text{obs}} - \boldsymbol{\mu}_{\text{model}} \right)^T \mathbf{C}^{-1} \left( \boldsymbol{\mu}_{\text{obs}} - \boldsymbol{\mu}_{\text{model}} \right)
\]

The components of this function include $\boldsymbol{\mu}_{\text{obs}}$, which encapsulates the distance moduli measurements from the observational dataset, while $\boldsymbol{\mu}_{\text{model}}$ represents the theoretical predictions derived from our cosmological framework. The matrix $\mathbf{C}^{-1}$ denotes the inverse of the covariance matrix, accounting for measurement uncertainties and their correlations. The posterior probability distribution $\mathcal{P}(\theta)$ is obtained through the product of this likelihood function and the chosen prior distributions.

\[
\mathcal{P}(\theta) \propto \mathcal{L}_{\text{SN}} \times \pi(\theta)
\]

\begin{table}[H]
\centering
\caption{Best-fit Parameters and Prior Ranges for BADE, NBADE, and $\Lambda$CDM Models}
\renewcommand{\arraystretch}{1.5}
\begin{tabular}{|l|l|c|c|c|}
\hline
\textbf{Model} & \textbf{Parameter} & \textbf{Pantheon+SH0ES} & \textbf{CC+BAO} & \textbf{Prior Range} \\
\hline
\multirow{5}{*}{BADE} 
& $H_{0}$ & $72.968_{-0.041}^{+0.023}$ & $69.059_{-0.903}^{+1.540}$ & $[60,\ 70]$ \\
& $\Omega_{m0}$ & $0.314_{-0.066}^{+0.034}$ & $0.326_{-0.060}^{+0.040}$ & $[0.1,\ 0.3]$ \\
& $\omega_m$ & $0.139_{-0.025}^{+0.036}$ & $0.005_{-0.029}^{+0.038}$ & $[-0.5,\ 0.5]$ \\
& $\Delta$ & $0.324_{-0.201}^{+0.322}$ & $0.389_{-0.275}^{+0.275}$ & $[0,\ 1]$ \\
& $M$ & $0.771_{-0.166}^{+0.163}$ & $0.714_{-0.136}^{+0.180}$ & $[-2,\ 2]$ \\
\hline
\multirow{5}{*}{NBADE} 
& $H_{0}$ & $72.992_{-0.014}^{+0.006}$ & $69.508_{-1.723}^{+2.348}$ & $[60,\ 72]$ \\
& $\Omega_{m0}$ & $0.309_{-0.002}^{+0.001}$ & $0.272_{-0.064}^{+0.066}$ & $[0.1,\ 0.3]$ \\
& $\omega_m$ & $0.132_{-0.014}^{+0.015}$ & $-0.028_{-0.043}^{+0.042}$ & $[-0.1,\ 1.2]$ \\
& $\Delta$ & $0.467_{-0.286}^{+0.351}$ & $0.451_{-0.243}^{+0.331}$ & $[0,\ 1]$ \\
& $M$ & $0.553_{-0.243}^{+0.270}$ & $0.626_{-0.301}^{+0.273}$ & $[0.09,\ 1.0]$ \\
\hline
\multirow{2}{*}{$\Lambda$CDM} 
& $H_{0}$ & $73.079_{-0.158}^{+0.157}$ & $69.989_{-1.197}^{+0.876}$ & $[60,\ 70]$ \\
& $\Omega_{m0}$ & $0.341_{-0.014}^{+0.006}$ & $0.268_{-0.015}^{+0.017}$ & $[0.1,\ 0.3]$ \\
\hline
\end{tabular}
\label{tab: best_fit_params_values}
\end{table}

\subsection{Information Criterion:}
For quantitative model evaluation, we employ three statistical metrics: the Akaike Information Criterion (AIC), the Bayesian Information Criterion (BIC), and the Deviance Information Criterion (DIC). The mathematical formulation of the AIC incorporates a finite sample size correction and is expressed as:

\[
\text{AIC} = -2 \ln(L_{\text{max}}) + 2k + \frac{2k(k+1)}{N_{\text{tot}} - k - 1},
\]

where the maximum likelihood value is denoted by $L_{\text{max}}$, the parameter count by $k$, and the total observational sample size by $N_{\text{tot}}$. The BIC framework provides an alternative statistical measure through the expression:

\[
\text{BIC} = -2 \ln(L_{\text{max}}) + k \ln(N_{\text{tot}}),
\]

maintaining consistent notation for the likelihood, parameter count, and sample size. The DIC methodology offers a hierarchical assessment through the relation:

\[
\text{DIC} = \bar{D} + p_D,
\]

where the mean posterior deviance $\bar{D}$ is computed by averaging over $S$ posterior samples:

\[
\bar{D} = \frac{1}{S} \sum_{i=1}^S D(\theta_i)
\]

The effective parameter count $p_D$ is determined via:

\[
p_D = \bar{D} - D(\hat{\theta}),
\]

with $D(\hat{\theta})$ representing the deviance evaluated at the posterior parameter means.

The comparative analysis between our proposed models and the standard $\Lambda$CDM framework is summarized in Table~\ref{tab:model_metrics_comparison}. For systematic comparison across multiple models, we compute the relative information criterion differences:

\[
\Delta IC_{\text{model}} = IC_{\text{model}} - IC_{\text{min}},
\]

where $IC_{\text{min}}$ represents the lowest criterion value among all considered models~\cite{anagnostopoulos2020observational}. The interpretation follows the Jeffreys scale~\cite{jeffreys1998theory}, which categorizes model compatibility as follows: $\Delta IC \leq 2$ indicates statistical consistency, $2 < \Delta IC < 6$ suggests moderate statistical tension, and $\Delta IC \geq 10$ signifies strong statistical disagreement between models.

\begin{table}[H]
\centering
\caption{Information Criterion and $\Delta$IC values for different models}
\begin{tabular}{|l|c|c|c|c|c|c|c|}
\hline
\textbf{Model} & \textbf{Criteria} & \textbf{AIC} & \textbf{BIC} & \textbf{DIC} & $\Delta$ \textbf{AIC} & $\Delta$ \textbf{BIC} & $\Delta$ \textbf{DIC} \\
\hline
\multicolumn{8}{|c|}{\textbf{Pantheon+SH0ES Dataset}} \\
\hline
BADE & Value & 1765.82 & 1776.02 & 1758.41 & 8.09 & 7.41 & 5.41 \\
NBADE & Value & 1758.76 & 1775.95 & 1753.00 & 1.03 & 7.34 & 0.00 \\
$\Lambda$CDM & Value & 1757.73 & 1768.61 & 1758.97 & 0.00 & 0.00 & 5.97 \\
\hline
\multicolumn{8}{|c|}{\textbf{CC+BAO Dataset}} \\
\hline
BADE & Value & 40.78 & 44.00 & 34.96 & 4.54 & 3.67 & 1.22 \\
NBADE & Value & 39.69 & 47.90 & 33.74 & 3.45 & 7.57 & 0.00 \\
$\Lambda$CDM & Value & 36.24 & 40.33 & 36.22 & 0.00 & 0.00 & 2.48 \\
\hline
\end{tabular}
\label{tab:model_metrics_comparison}
\end{table}

\section{Results and Discussion} \label{Sect: Result}
{\bf Comparison of Hubble Parameter Plot and Relative Difference:} The Hubble parameter values for the BADE and NBADE models are as follows: for BADE, $H_0 = 72.968^{+0.023}_{-0.041}$ km s$^{-1}$ Mpc$^{-1}$ (Pantheon+SH0ES) and $H_0 = 69.059^{+1.540}_{-0.903}$ km s$^{-1}$ Mpc$^{-1}$ (CC+BAO); for NBADE, $H_0 = 72.992^{+0.006}_{-0.014}$ km s$^{-1}$ Mpc$^{-1}$ (Pantheon+SH0ES) and $H_0 = 69.508^{+2.348}_{-1.723}$ km s$^{-1}$ Mpc$^{-1}$ (CC+BAO). These values are in alignment with the estimates obtained from the Planck Collaboration \cite{refId0}, while they exhibit a very slight deviation from the value obtained from the SH0ES collaboration ($73.24\pm 1.74$ km s$^{-1}$ Mpc$^{-1}$) in 2019 \cite{riess2019large} but still very close to it. 

\begin{figure}[htbp]
    \centering
    \begin{subfigure}[b]{0.45\textwidth}
        \includegraphics[width=\textwidth]{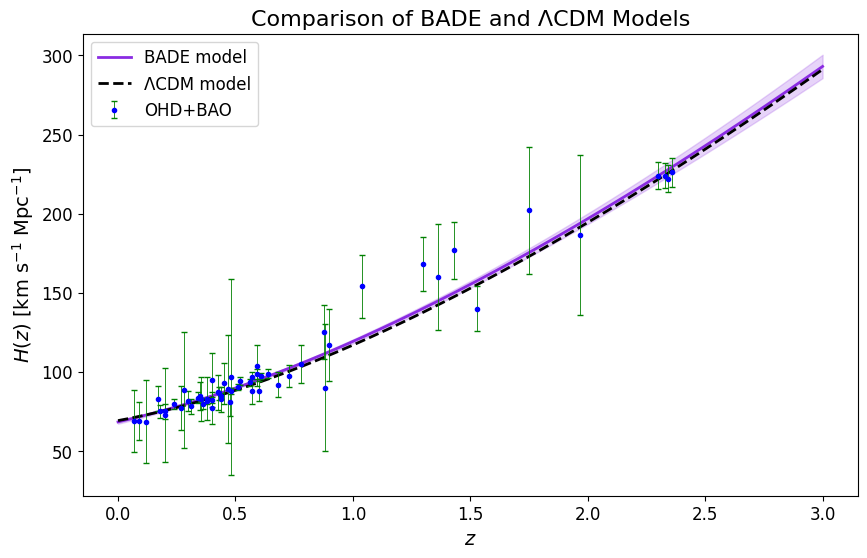}
        \caption{}
        \label{fig:Hubble_BADE}
    \end{subfigure}
    \begin{subfigure}[b]{0.45\textwidth}
        \includegraphics[width=\textwidth]{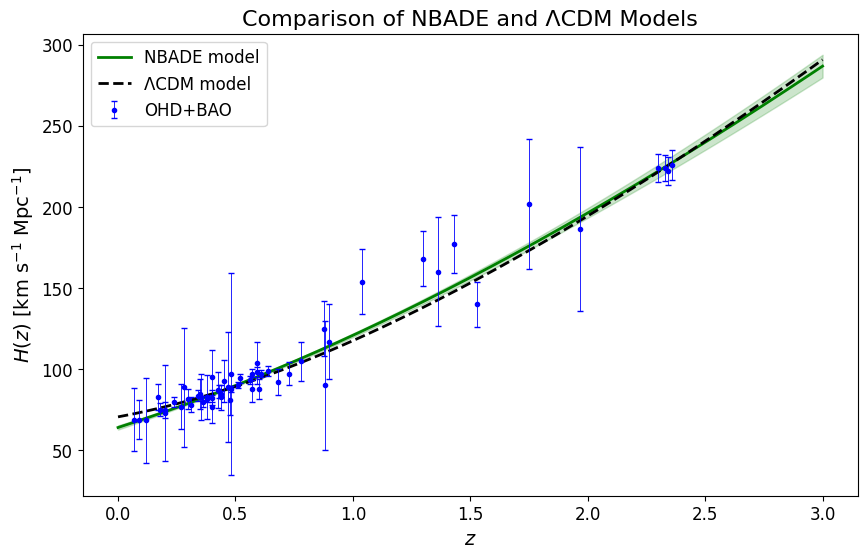}
        \caption{}
        \label{fig:Hubble_NBADE}
    \end{subfigure}
    \caption{Comparison of Hubble Parameter between $\Lambda$CDM and the fig: (a) BADE model and fig: (b) NBADE model using 57 CC + BAO data points.}
    \label{fig:Hubble_BADE_NBADE}
\end{figure}

The redshift evolution of the Hubble parameter $H(z)$ is visualized in Fig.~\ref{fig:Hubble_BADE_NBADE}, where we contrast our reconstructed Einstein-Aether gravity solutions (BADE and NBADE models) with the standard $\Lambda$CDM cosmology. The observational constraints from the CC+BAO dataset, depicted as blue data points, provide empirical anchors for our theoretical frameworks. The comparative analysis, detailed in Fig.~\ref{fig:Hubble_BADE} and Fig.~\ref{fig:Hubble_NBADE}, demonstrates remarkable concordance between theoretical predictions and observational measurements, validating our models' capacity to characterize cosmic evolutionary dynamics.

Our analysis includes the benchmark $\Lambda$CDM model, parameterized by the conventional matter density $\Omega_{m0} = 0.3$ and dark energy density $\Omega_{d0} = 0.7$. The visualization in Fig.~\ref{fig:Hubble_BADE_NBADE} reveals that while both BADE and NBADE frameworks exhibit strong consistency with $\Lambda$CDM predictions at proximate cosmic distances, they manifest distinct evolutionary trajectories at higher redshifts, introducing subtle yet noteworthy deviations from the standard cosmological paradigm.

\begin{figure}[htbp]
    \centering
    \begin{subfigure}[b]{0.45\textwidth}
        \includegraphics[width=\textwidth]{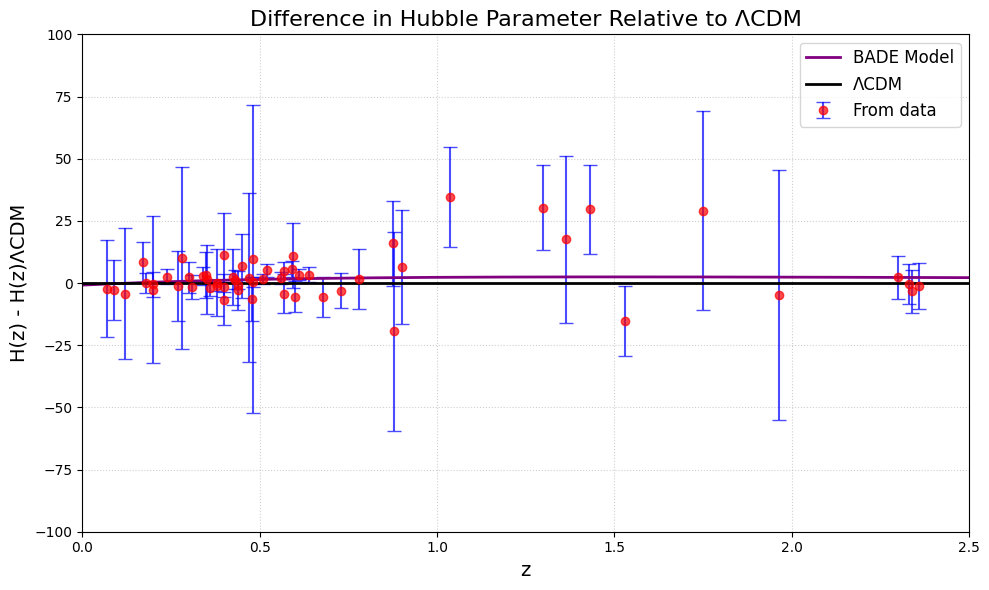}
        \caption{}
        \label{fig:Relative Hubble_BADE}
    \end{subfigure}
    \begin{subfigure}[b]{0.45\textwidth}
        \includegraphics[width=\textwidth]{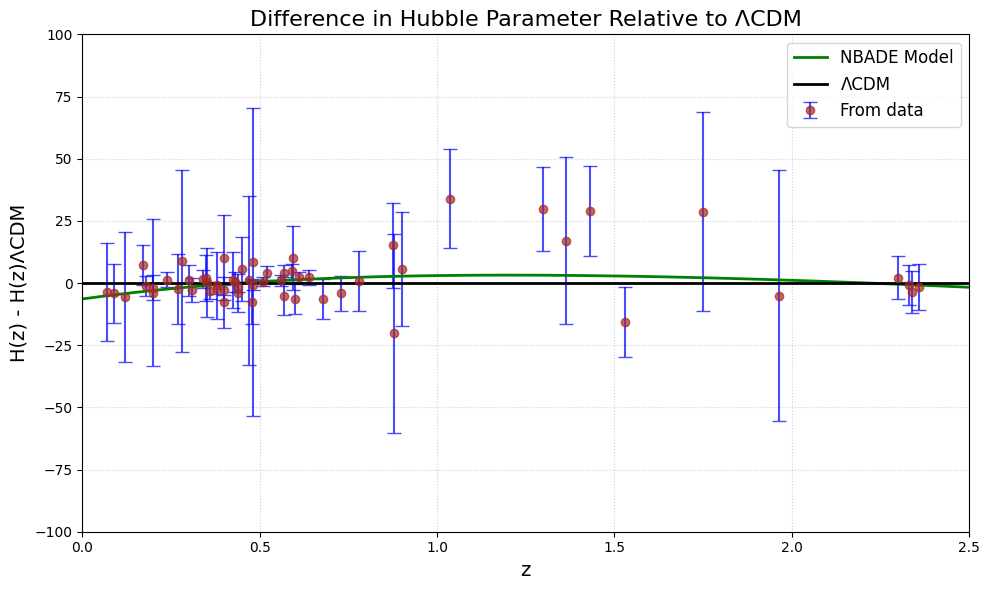}
        \caption{}
        \label{fig:Relative Hubble_NBADE}
    \end{subfigure}
    \caption{Comparison of variation of difference in Hubble Parameter between $\Lambda$CDM and the fig: (a) BADE model and fig: (b) NBADE model gravity using 57 CC + BAO data points.}
    \label{fig:Relative_Hubble_BADE_NBADE}
\end{figure}

A quantitative assessment of the model discrepancies is presented in Fig.~\ref{fig:Relative_Hubble_BADE_NBADE}, which examines the differential evolution between our proposed frameworks and the standard $\Lambda$CDM cosmology. The analysis reveals a characteristic pattern: at intermediate to high redshifts $(z > 0.5)$, both theoretical constructs exhibit measurable departures from $\Lambda$CDM predictions when evaluated against the cosmic chronometer (CC) and BAO observational datasets. This systematic divergence at elevated redshifts suggests the presence of distinct physical mechanisms or parametric dependencies that become increasingly prominent in these cosmological regimes.

The behavior in the low-redshift domain $(z < 0.5)$ presents a contrasting scenario, where the theoretical predictions demonstrate progressive convergence. Both BADE and NBADE formulations exhibit diminishing deviations from the $\Lambda$CDM framework as the redshift approaches contemporary epochs. This convergent behavior at proximate cosmic distances is particularly significant, as it establishes consistency with the high-precision measurements derived from cosmic chronometers and Baryon Acoustic Oscillation surveys in the nearby universe.

\begin{figure}[htbp]
    \centering
    \begin{subfigure}[b]{0.45\textwidth}
        \includegraphics[width=\textwidth]{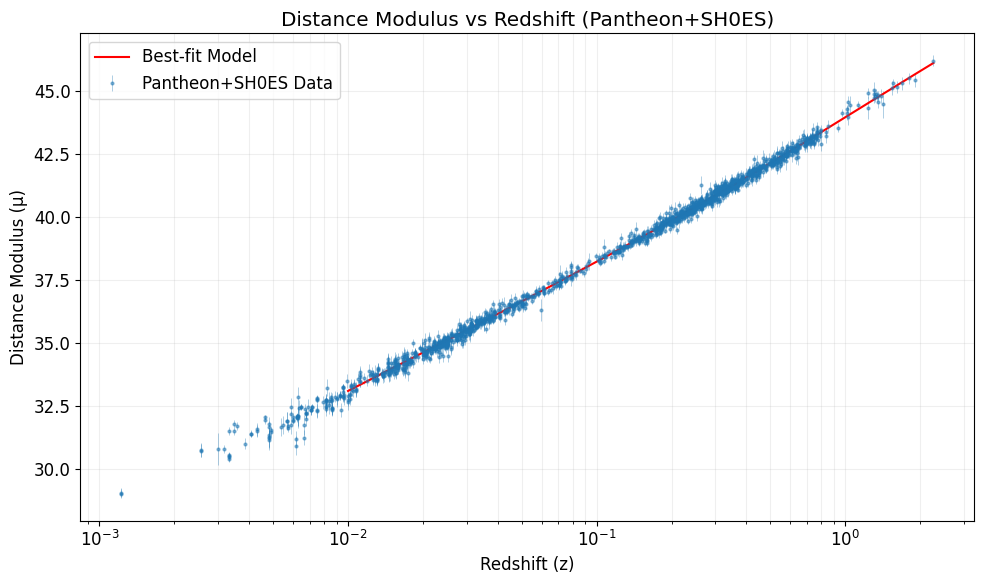}
        \caption{}
        \label{fig:Distance_modulus_BADE}
    \end{subfigure}
    \begin{subfigure}[b]{0.45\textwidth}
        \includegraphics[width=\textwidth]{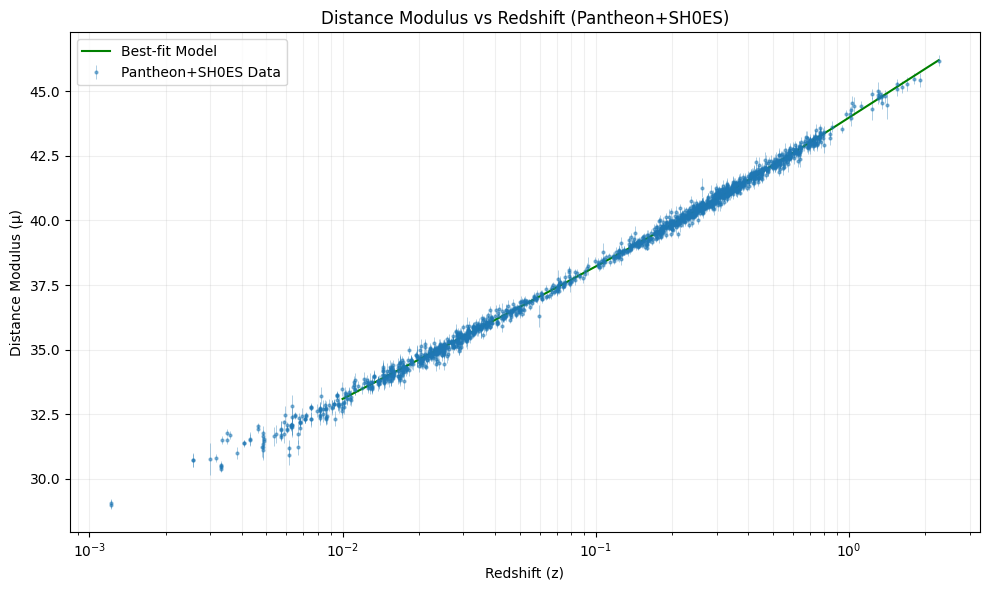}
        \caption{}
        \label{fig:Distance_modulus_NBADE}
    \end{subfigure}
    \caption{Best fit plot of Distance modulus against redshift for fig: (a) BADE model and fig: (b) NBADE model gravity using 1701 Pantheon+SH0ES data points.}
    \label{fig:Distance_modulus_BADE_NBADE}
\end{figure}

{\bf Distance Modulus Comparison:} Figure \ref{fig:Distance_modulus_BADE_NBADE} displays the distance modulus \((\mu)\) as a function of redshift \((z)\) for both the BADE and NBADE models, fitted to the Pantheon+SH0ES dataset. The observed data points are plotted as blue dots with error bars, while the best-fit models are represented by solid lines (red for BADE in fig.~\ref{fig:Distance_modulus_BADE}, green for NBADE in fig.~\ref{fig:Distance_modulus_NBADE}). Both models demonstrate excellent agreement with the observational data across the entire redshift range ($10^{-3} < z < 2.3$). The distance modulus increases monotonically with redshift, as expected in an expanding universe. This relationship reflects the growing distance light must travel from more distant (higher redshift) sources, resulting in fainter apparent magnitudes and, thus, larger distance moduli. 

At low redshifts ($z < 0.01$), both models accurately capture the near-linear relationship between distance modulus and redshift (on a log scale). This region is crucial for constraining the Hubble constant ($H_0$). In the range $0.01 < z < 0.1$, we observe a smooth transition in the curve's slope, indicating the increasing influence of cosmic expansion on the distance-redshift relation. In the range $0.01 < z < 0.1$, we observe a smooth transition in the curve's slope, indicating the increasing influence of cosmic expansion on the distance-redshift relation. The close agreement between both models and the observational data indicates that the BADE and NBADE frameworks effectively describe the evolution history of the universe as evidenced by Type Ia supernovae.

\begin{figure}[htbp]
    \centering
    \begin{subfigure}[b]{0.45\textwidth}
        \includegraphics[width=\textwidth]{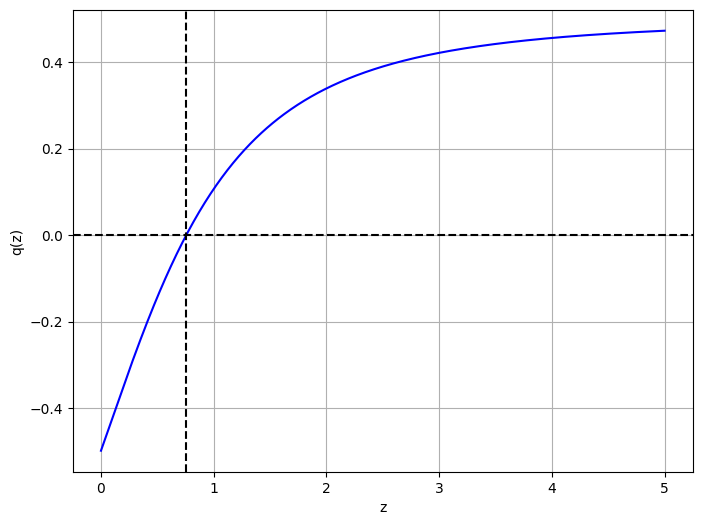}
        \caption{}
        \label{fig:q_BADE}
    \end{subfigure}
    \begin{subfigure}[b]{0.45\textwidth}
        \includegraphics[width=\textwidth]{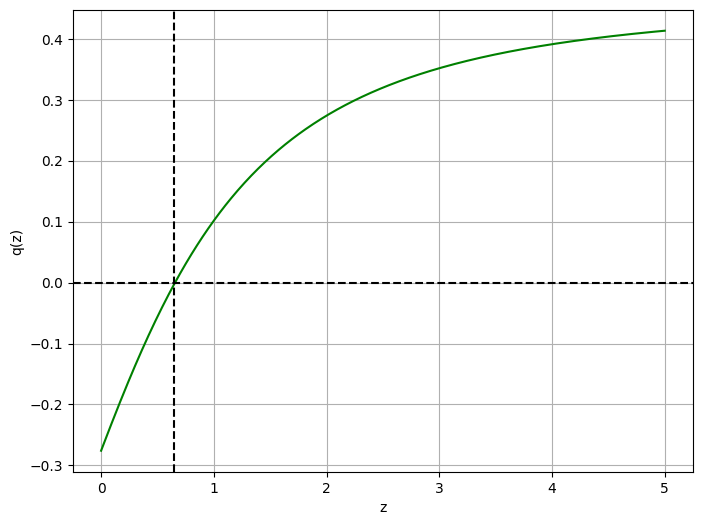}
        \caption{}
        \label{fig:q_NBADE}
    \end{subfigure}
    \caption{Redshift dependence of the universal deceleration parameter illustrated for: (a) the BADE cosmological framework and (b) the NBADE theoretical construction.}
    \label{fig:q_BADE_NBADE}
\end{figure}

{\bf Cosmographic Parameters:} The temporal evolution of the cosmic deceleration parameter $q(z)$ is illustrated in Fig.~\ref{fig:q_BADE_NBADE}, revealing the critical transition epochs between decelerating and accelerating phases of universal expansion. The analysis identifies distinctive transition redshifts: $z_t \approx 0.7568$ characterizes the BADE framework, while $z_t \approx 0.6561$ marks the transition point in the NBADE formulation. These determinations demonstrate compatibility with the empirical constraints established through combined analyses of Type Ia supernovae, cosmic microwave background radiation, and large-scale structure observations.

The differential positioning of these transition points, notably the elevated $z_t$ value in the NBADE scenario, indicates fundamental distinctions in the acceleration mechanisms between the two theoretical frameworks. Contemporary evaluations of the deceleration parameter, measured at the present epoch $(z=0)$, yield characteristically different values: the BADE model predicts $q_0 = -0.4985$, while the NBADE formulation suggests a more moderate deceleration with $q_0 = -0.2759$.

\begin{figure}[htbp]
    \centering
    \begin{subfigure}[b]{0.45\textwidth}
        \includegraphics[width=\textwidth]{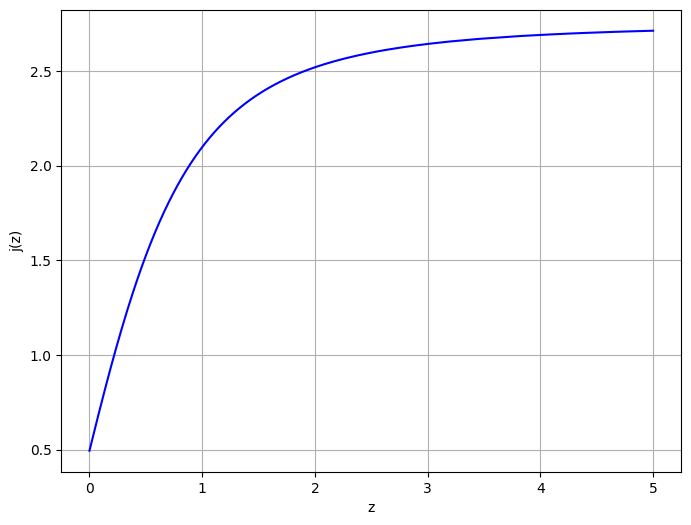}
        \caption{}
        \label{fig:j_BADE}
    \end{subfigure}
    \begin{subfigure}[b]{0.45\textwidth}
        \includegraphics[width=\textwidth]{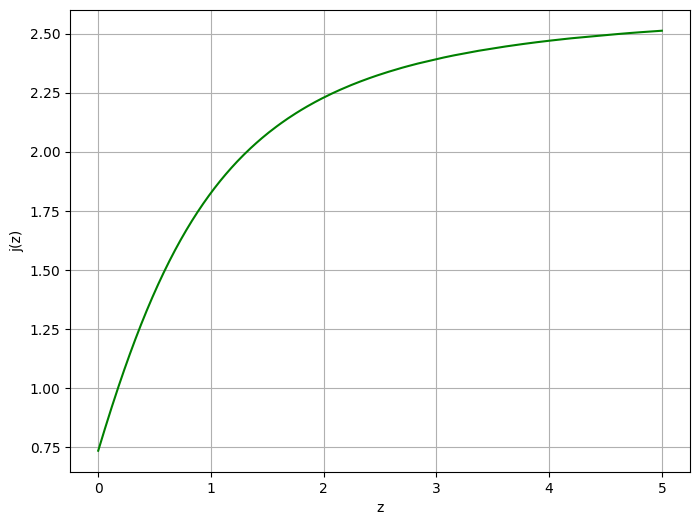}
        \caption{}
        \label{fig:j_NBADE}
    \end{subfigure}
    \caption{Redshift dependence of the universal jerk parameter illustrated for: (a) the BADE cosmological framework and (b) the NBADE theoretical construction.}
    \label{fig:j_BADE_NBADE}
\end{figure}

\begin{figure}[htbp]
    \centering
    \begin{subfigure}[b]{0.45\textwidth}
        \includegraphics[width=\textwidth]{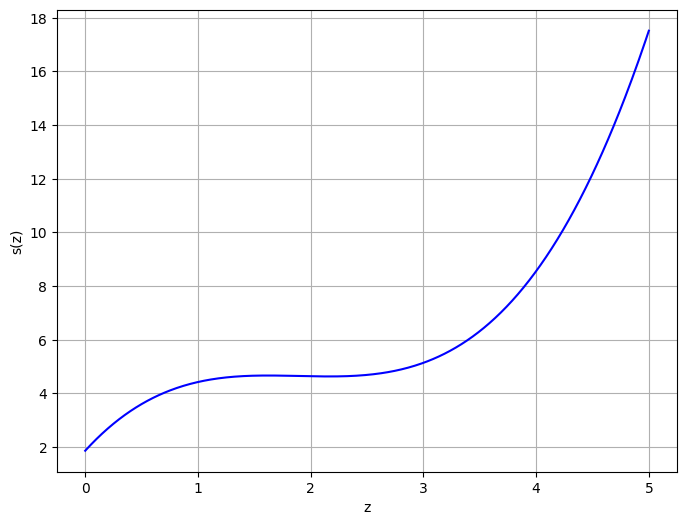}
        \caption{}
        \label{fig:s_BADE}
    \end{subfigure}
    \begin{subfigure}[b]{0.45\textwidth}
        \includegraphics[width=\textwidth]{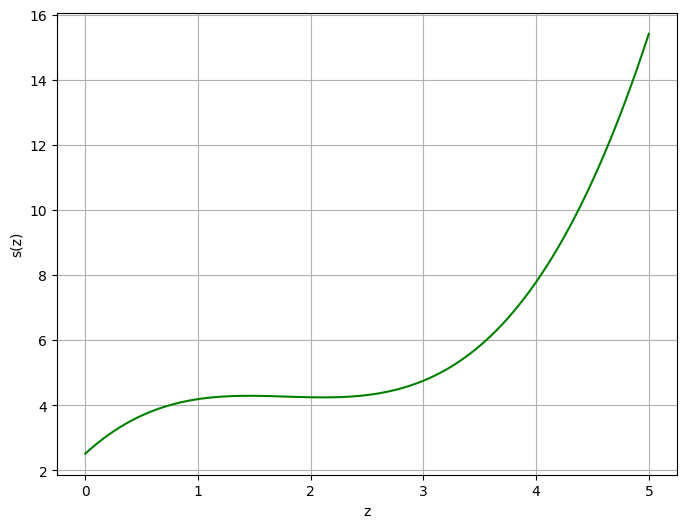}
        \caption{}
        \label{fig:s_NBADE}
    \end{subfigure}
    \caption{Redshift dependence of the universal snap parameter illustrated for: (a) the BADE cosmological framework and (b) the NBADE theoretical construction.}
    \label{fig:s_BADE_NBADE}
\end{figure}
\begin{figure}[htbp]
    \centering
    \begin{subfigure}[b]{0.45\textwidth}
        \includegraphics[width=\textwidth]{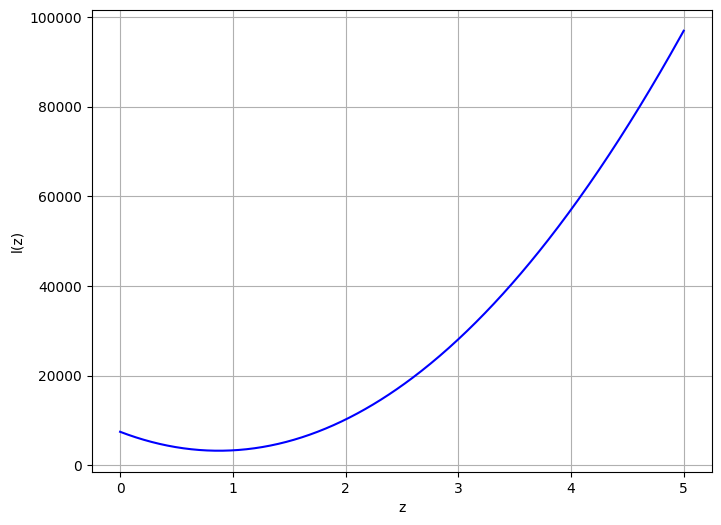}
        \caption{}
        \label{fig:l_BADE}
    \end{subfigure}
    \begin{subfigure}[b]{0.45\textwidth}
        \includegraphics[width=\textwidth]{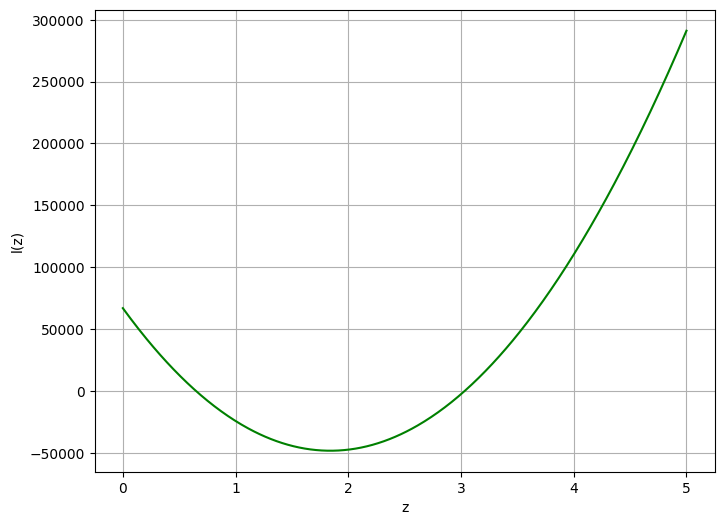}
        \caption{}
        \label{fig:l_NBADE}
    \end{subfigure}
    \caption{Redshift dependence of the universal lerk parameter illustrated for: (a) the BADE cosmological framework and (b) the NBADE theoretical construction.}
    \label{fig:l_BADE_NBADE}
\end{figure}

In Fig.~\ref{fig:j_BADE_NBADE}, the jerk parameter is shown to have positive values throughout the evolution for both models, indicating that the accelerated phase of the universe's expansion is currently increasing. The snap and lerk parameters, depicted in Figs.~\ref{fig:s_BADE} and \ref{fig:l_BADE} for the BADE model, exhibit positive values across the entire redshift range. In Figs.~\ref{fig:s_NBADE} and \ref{fig:l_NBADE} for the NBADE model, while the snap parameter remains positive, the lerk parameter becomes negative at certain redshift intervals before returning to positive for $z < 1$. This behavior indicates that both the jerk and snap parameters are increasing over time. This trend indicates that, over time, the rate of acceleration of the universe's expansion is increasing, accompanied by a growing rate of change in that acceleration.

\begin{figure}[htbp]
    \centering
    \begin{subfigure}[b]{0.45\textwidth}
        \includegraphics[width=\textwidth]{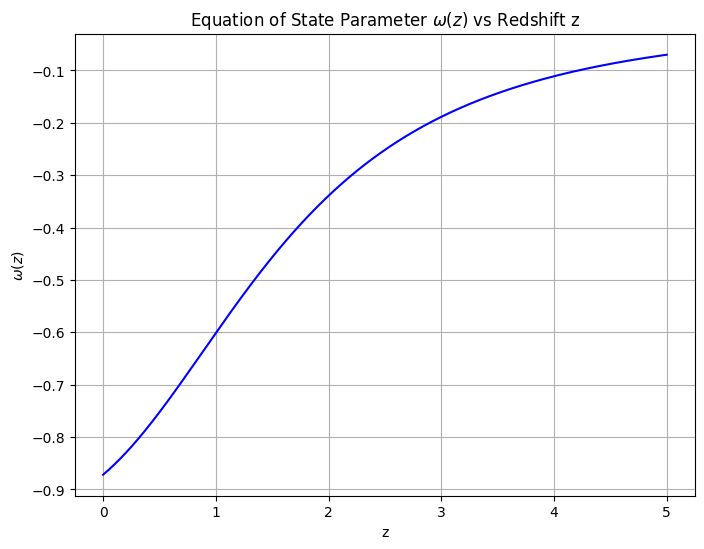}
        \caption{}
        \label{fig:omega_BADE}
    \end{subfigure}
    \begin{subfigure}[b]{0.45\textwidth}
        \includegraphics[width=\textwidth]{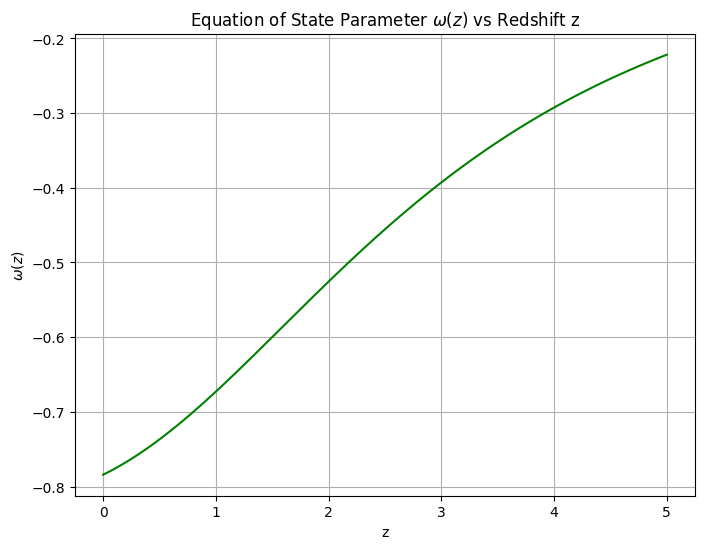}
        \caption{}
        \label{fig:omega_NBADE}
    \end{subfigure}
    \begin{subfigure}[b]{0.55\textwidth}
        \includegraphics[width=\textwidth]{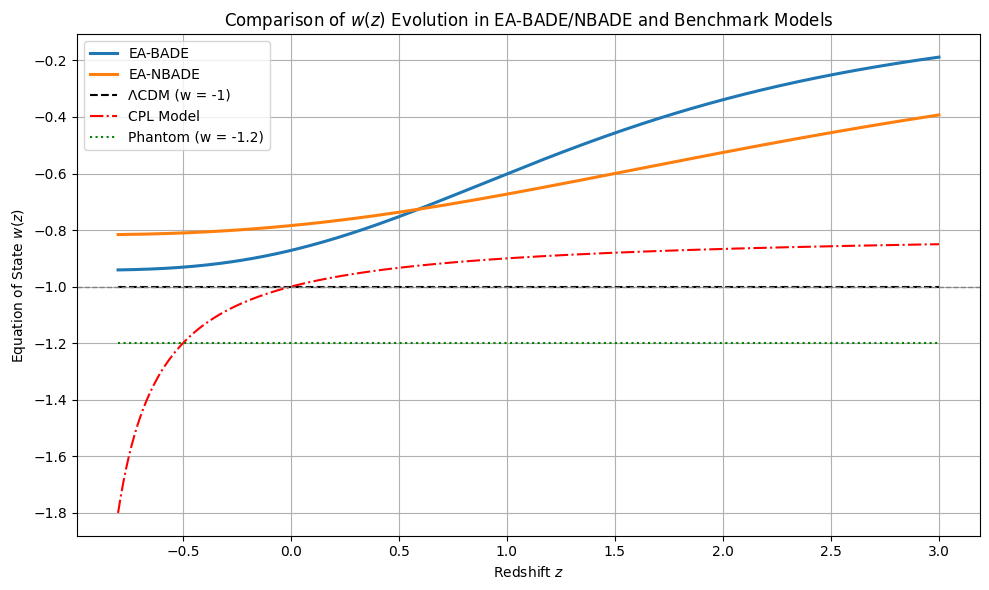}
        \caption{}
        \label{fig:omega_NBADE_BADE_Comparison}
    \end{subfigure}
    \caption{Plot of evolution trajectory of EoS parameter for dark energy $\omega_{DE}(z)$ for the fig: (a) BADE model and fig: (b) NBADE model. (c) A precise comparison with other models.}
    \label{fig:omega_BADE_NBADE}
\end{figure}

\begin{figure}[htbp]
    \centering
    \begin{subfigure}[b]{0.45\textwidth}
        \includegraphics[width=\textwidth]{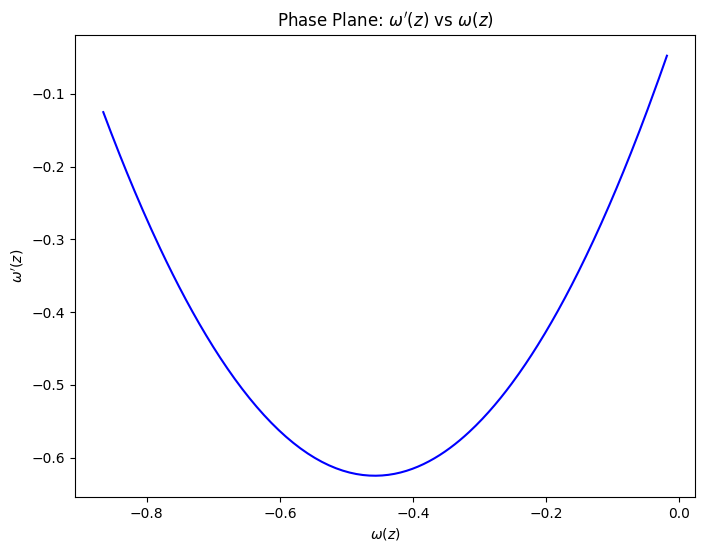}
        \caption{}
        \label{fig:omega'_BADE}
    \end{subfigure}
    \begin{subfigure}[b]{0.45\textwidth}
        \includegraphics[width=\textwidth]{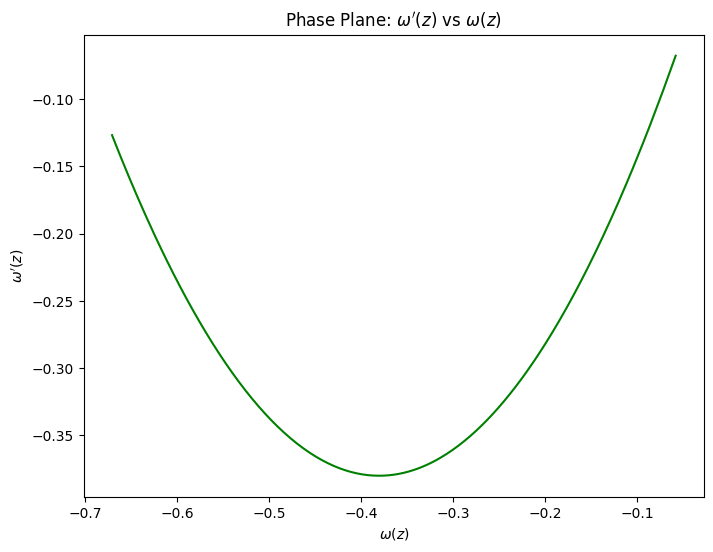}
        \caption{}
        \label{fig:omega'_NBADE}
    \end{subfigure}
    \caption{Plot of $\omega'_{DE}-\omega_{DE}$ phase plane for the fig: (a) BADE model and fig: (b) NBADE model.}
    \label{fig:omega'_BADE_NBADE}
\end{figure}

{\bf EoS Parameter and the $\omega'_{DE}-\omega_{DE}$ Phase Plane:} The dynamical characteristics of the dark energy equation of state (EoS) parameter $\omega_{DE}$ are presented in Fig.~\ref{fig:omega_BADE_NBADE}. The analysis reveals a consistently negative parametric trajectory throughout cosmic history, with an upward trend as the universe approaches the present epoch (decreasing $z$). This characteristic pattern indicates the emergence and subsequent dominance of the dark energy component. The observed behavior in both theoretical frameworks - BADE and NBADE - exhibits signatures consistent with quintessence-like dark energy dynamics. Contemporary measurements yield distinct present-day values: the BADE construction predicts $\omega_{DE0} = -0.8718$, while the NBADE formalism suggests $\omega_{DE0} = -0.7841$. These theoretical predictions demonstrate close alignments from contemporary observational constraints from DES-SN5YR \cite{Abbott_2024} and DESI-VI \cite{Adame_2025}. In other words, the models are in the quintessence regime today. However, both $\omega_{DE}(z)$ curves move towards $-1$ as redshift decreases, indicating an asymptotic crossing into the phantom regime ($\omega_{DE}<-1$) in the future. This crossing of the $w=-1$ divide is not abrupt but happens at $z<0$ (i.e. it lies in the extrapolated future). Such behavior is not unique to the Einstein-Aether reconstructions: it is well known, for example, that Barrow holographic dark energy can similarly evolve from quintessence to phantom as parameters change~\cite{Pradhan_2021}. Indeed, many modified gravity or interacting dark energy scenarios can achieve an effective phantom crossing without pathologies~\cite{Caldwell_2002}. In contrast, a simple canonical quintessence model has $\omega_{DE}>-1$ at all times, and models like the generalized Chaplygin gas smoothly approach $\omega_{DE}\to-1$ from above (never truly crossing) at late times. Thus our models’ eventual phantom-like fate (driven by the Barrow exponent effects and Aether coupling) is a novel prediction, but it parallels trends seen in other exotic DE frameworks.
The phase-plane trajectory in the $(\omega_{DE},\omega'_{DE})$ space (Fig.~\ref{fig:omega'_BADE_NBADE}) lies in the so-called freezing region: the curves move toward $\omega_{DE}=-1$ with a slowly decreasing derivative (i.e.\ $\omega'_{DE}<0$). This indicates that dark energy is dynamically settling into a cosmological constant–like state as time passes. Freezing behavior is characteristic of many tracking quintessence models~\cite{Caldwell_2005} and differs from “thawing” models where $w$ evolves away from $-1$. Thus, even though our present-day $\omega_{DE}$ is slightly above $-1$, its future trend is toward $-1$, implying ongoing acceleration that will asymptotically resemble de Sitter expansion.
 
\begin{figure}[htbp]
    \centering
    \begin{subfigure}[b]{0.45\textwidth}
        \includegraphics[width=\textwidth]{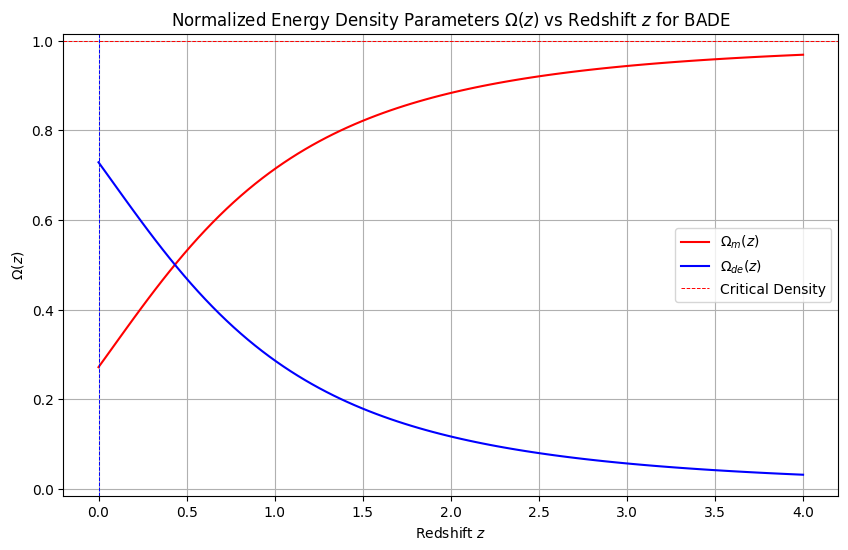}
        \caption{}
        \label{fig:Omega_total_BADE}
    \end{subfigure}
    \begin{subfigure}[b]{0.45\textwidth}
        \includegraphics[width=\textwidth]{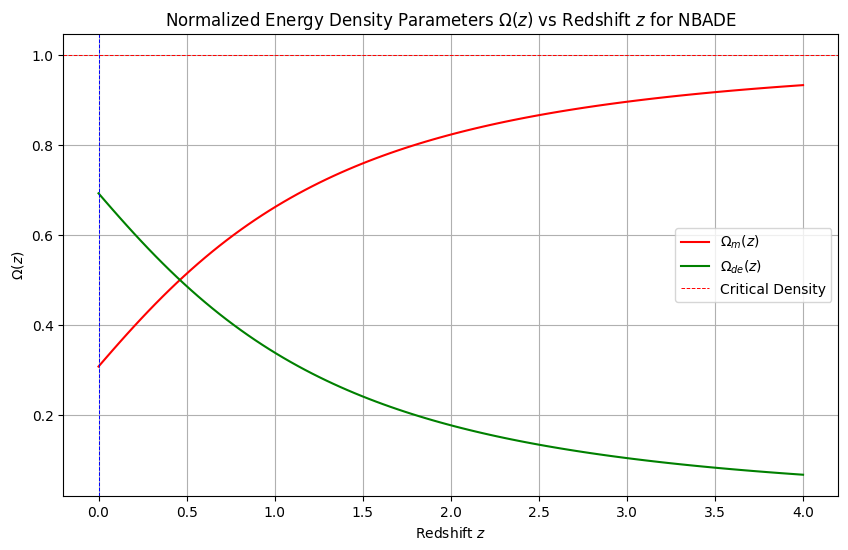}
        \caption{}
        \label{fig:Omega_total_NBADE}
    \end{subfigure}
    \caption{Plot of density parameter components of matter and dark energy for the fig: (a) BADE model and fig: (b) NBADE model.}
    \label{fig:Omega_total_BADE_NBADE}
\end{figure}

{\bf Density Parameter:} The temporal progression of the normalized density parameters $\Omega(z)$ can be observed in Figure~\ref{fig:Omega_total_BADE_NBADE}, which presents a comparative analysis between BADE and NBADE cosmological frameworks. The visualization emphasizes two key cosmological constituents: the matter component ($\Omega_m(z)$) and the dark energy contribution ($\Omega_{de}(z)$). At $z=0$, denoted by a vertical crimson line, we observe the critical density threshold. Analysis of the temporal evolution reveals that $\Omega_m(z)$ exhibits an ascending trend toward higher redshifts, demonstrating the pronounced role of matter in the early cosmic epochs. In contrast, $\Omega_{de}(z)$ displays an inverse relationship with redshift, indicating the growing prominence of dark energy in recent cosmic history. A significant transition point emerges at $z \approx 0.5$, where these components achieve parity, marking the shift between matter and dark energy dominance. The analysis confirms the preservation of cosmic flatness, evidenced by the constant sum $\Omega_m + \Omega_{de} = 1$ across all epochs. This relationship validates the fundamental assumption of spatial flatness in our cosmological model.

\begin{figure}[htbp]
    \centering
    \begin{subfigure}[b]{0.45\textwidth}
        \includegraphics[width=\textwidth]{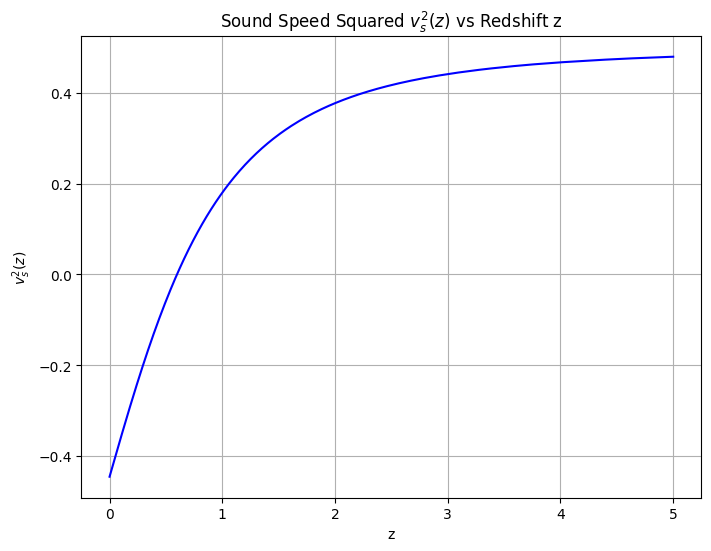}
        \caption{}
        \label{fig:v^2_BADE}
    \end{subfigure}
    \begin{subfigure}[b]{0.45\textwidth}
        \includegraphics[width=\textwidth]{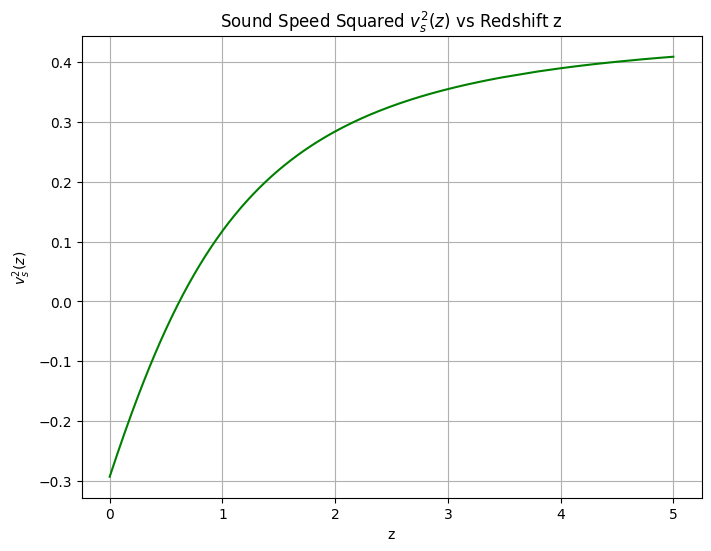}
        \caption{}
        \label{fig:v^2_NBADE}
    \end{subfigure}
    \caption{Graph depicting the square of the speed of sound \(v_s^2\) for (a) the BADE model and (b) the NBADE model.}
    \label{fig:v^2_BADE_NBADE}
\end{figure}

{\bf Square of the Speed of Sound:} The models’ viability as cosmological fluids can be assessed via the squared sound speed $v_s^2$ (Fig.~\ref{fig:v^2_BADE_NBADE}). For both EA-BADE and EA-NBADE, $v_s^2$ is positive at high redshift and becomes negative at late times, indicating that the reconstructed Aether–DE mixture is stable (no perturbation growth) in the past but develops an instability in the recent universe. In other words, both models are partially stable: small early perturbations do not grow, but late-time perturbations could. This echoes findings in some agegraphic or holographic models where stability is marginal. This behavior is consistent with recent studies on agegraphic and holographic dark energy models, which report similar partial stability, with stability at high redshifts and marginal or unstable conditions at late times in modified gravity frameworks \cite{SHARIF2024101606, KUMAR2024102085, Ghaffari_2022, Linton_2018}. In contrast, Ajmal and Sharif (2025)~\cite{ajmal2025newagegraphicdarkenergy} found complete instability $(v_s^2>0)$ throughout cosmic evolution in a new agegraphic dark energy model within modified symmetric teleparallel theory, while Abdollahi Zadeh et al. (2018)~\cite{Zadeh_2018} reported persistent instability in ghost dark energy within the DGP braneworld, regardless of interaction scenarios and also Liu et al. (2020)~\cite{liu2020interacting} reported the same for interacting ghost dark energy in a complex quintessence background. These comparisons underscore the distinct stability profile of EA-BADE and EA-NBADE within Agegraphic Dark Energy framework.

\begin{figure}[htbp]
    \centering
    \begin{subfigure}[b]{0.45\textwidth}
        \includegraphics[width=\textwidth]{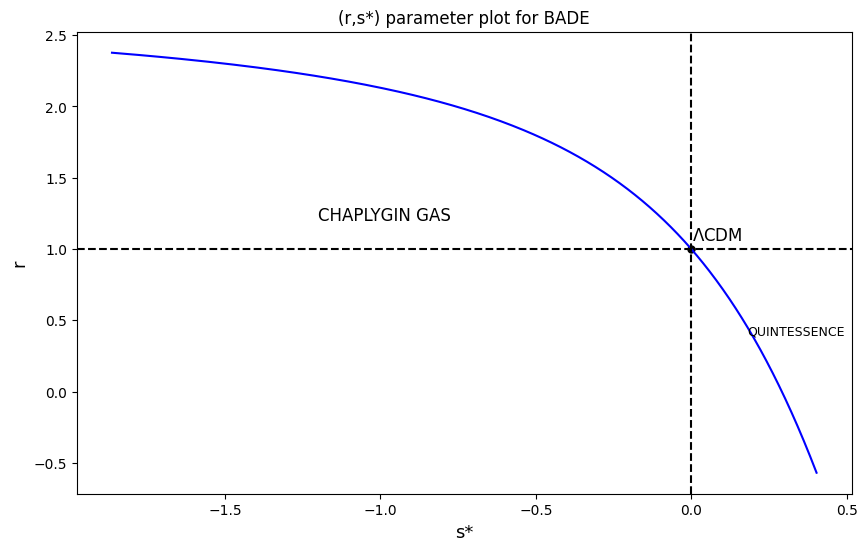}
        \caption{}
        \label{fig:r_s_BADE}
    \end{subfigure}
    \begin{subfigure}[b]{0.45\textwidth}
        \includegraphics[width=\textwidth]{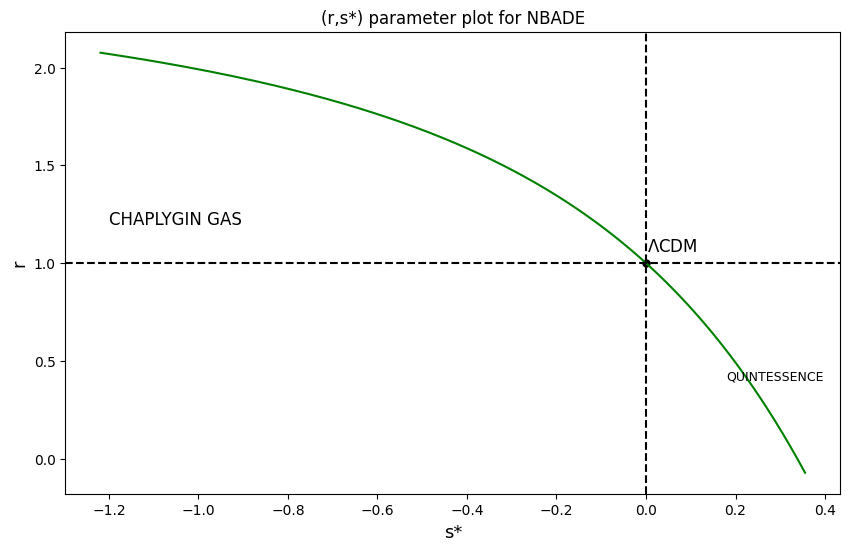}
        \caption{}
        \label{fig:r_s_NBADE}
    \end{subfigure}
    \caption{Plot of evolution trajectory of $(r,s^*)$ statefinder parameter for the fig: (a) BADE model and fig: (b) NBADE model for reconstructed Einstein-Aether gravity.}
    \label{fig:r_s_BADE_NBADE}
\end{figure}

\begin{figure}[htbp]
    \centering
    \begin{subfigure}[b]{0.45\textwidth}
        \includegraphics[width=\textwidth]{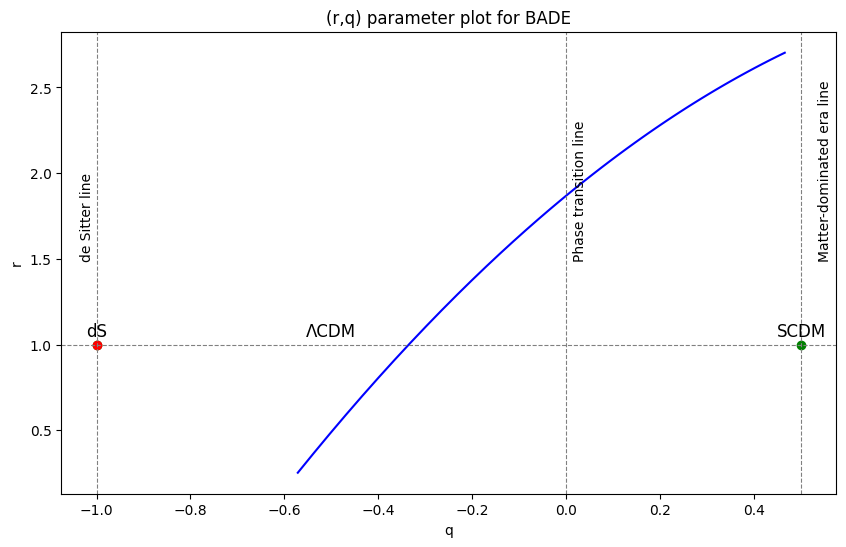}
        \caption{}
        \label{fig:r_q_BADE}
    \end{subfigure}
    \begin{subfigure}[b]{0.45\textwidth}
        \includegraphics[width=\textwidth]{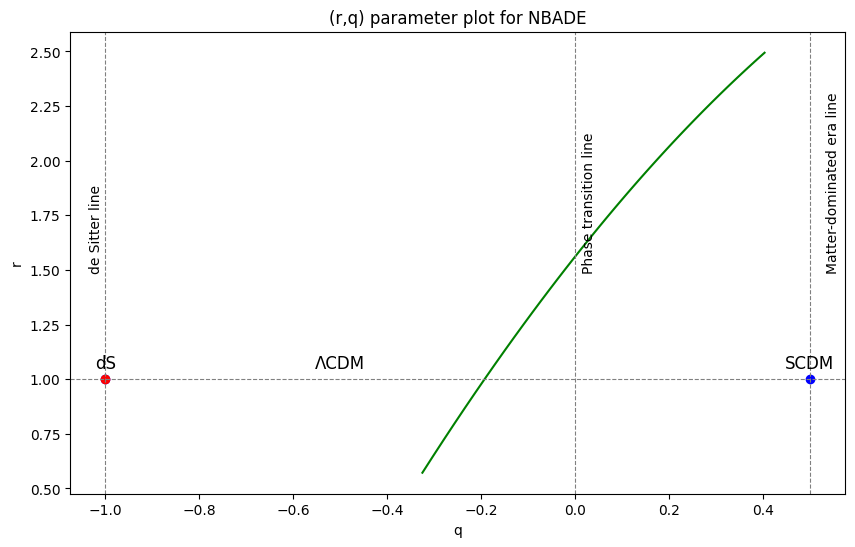}
        \caption{}
        \label{fig:r_q_NBADE}
    \end{subfigure}
    \caption{Plot of evolution trajectory of $(r,q)$ statefinder parameter for the fig: (a) BADE model and fig: (b) NBADE model for reconstructed Einstein-Aether gravity.}
    \label{fig:r_q_BADE_NBADE}
\end{figure}

{\bf Statefinder Parameter:} The analysis of dynamical parameters in Figure~\ref{fig:r_s_BADE_NBADE} reveals the evolutionary trajectory in the \((r, s^*)\) plane. During early cosmic epochs, both cosmological frameworks exhibit characteristics defined by \((r > 1)\) and \((s^* < 0)\), corresponding to quintessence behavior. The evolutionary path intersects the \(\Lambda\)CDM fixed point at \((r, s^*) = (1, 0)\), subsequently entering a domain characterized by \((r < 1)\) and \((s^* > 0)\), which is indicative of Chaplygin gas (Phantom-like) behavior. The dynamical evolution portrayed in Figure~\ref{fig:r_q_BADE_NBADE} through the \((r, q)\) plane demonstrates the cosmic transition from decelerated to accelerated expansion. This transformation reflects the shift from matter dominance to dark energy prevalence. Notable is the absence of convergence to either the steady-state configuration \((r = 1, q = -1)\) or the SCDM point \((r = 1, q = \frac{1}{2})\). Instead, the asymptotic behavior of both models tends toward the de Sitter state, characterized by \(q = -1\). In the context of Einstein-Aether gravity reconstruction, the BADE framework exhibits more rapid convergence compared to its NBADE counterpart. These behaviors align with recent studies on statefinder diagnostics in dark energy and modified gravity models, particularly in Barrow agegraphic and Einstein-Aether frameworks, which report similar quintessence-to-phantom transitions and de Sitter convergence in the \((r, s^*)\) and \((r, q)\) planes \cite{Tamal_f(P)_f(Q), Kumar2023, Srivastava2021, Singh2016, Carrasco2024}.

An important dynamical feature observed in both reconstructed Einstein-Aether gravity models—corresponding to BADE and NBADE—is the apparent transition of the effective equation of state (EoS) parameter $\omega_{DE}$ from a quintessence regime ($\omega_{DE} > -1$) at present to a phantom regime ($\omega_{DE} < -1$) in the future. While such a transition is a known characteristic of various dynamical dark energy (DDE) models \cite{Tamal_f(P)_f(Q),debnath2014reconstructions}, it is crucial to determine whether this behavior in our framework exhibits any model-specific features—such as a distinct transition redshift, the sharpness or smoothness of the crossing, or dependence on model parameters—that may offer discriminative power against other scenarios. In our reconstructed EA-BADE model, the EoS parameter smoothly evolves from $\omega_{DE} \approx -0.86$ at $z = 0$ to $\omega_{DE} \approx -1.05$ in the far future ($z \to -1$). For the EA-NBADE model, the present value is slightly closer to the phantom divide with $\omega_{DE} \approx -0.91$ at $z = 0$, evolving to $\omega_{DE} \approx -1.12$ as $z \to -1$. Interestingly, although other diagnostic tools—such as the $(r, s)$ statefinder trajectories and the $\text{Om}(z)$ plots—suggest a late-time shift toward phantom-like behavior, the direct evolution of $\omega_{DE}(z)$ in both models does \textit{not} clearly exhibit a definitive crossing of the phantom divide within the accessible redshift range. This apparent tension can be interpreted as a subtle signal of an \textit{asymptotic approach} toward the phantom regime rather than a sharp transition. In particular, the value of $\omega_{DE}$ decreases smoothly and increasingly deviates from $-1$ in the negative redshift direction, but without showing a pronounced crossing point—suggesting that the models may mimic phantom behavior in diagnostics without realizing a full quintessence-to-phantom transition in the strict sense. This comparative behavior is illustrated in Figure~\ref{fig:omega_NBADE_BADE_Comparison}, where we have plotted $\omega_{DE}(z)$ for EA-BADE and EA-NBADE alongside benchmark models such as CPL and Holographic Dark Energy. Notably, in parameter space, the proximity to the phantom regime is influenced by the Barrow exponent $\Delta$, which modulates the evolution rate of the agegraphic energy density. Larger $\Delta$ values tend to delay the approach to the phantom-like regime and yield a smoother evolution of $\omega_{DE}$, a feature that can potentially be tested with upcoming precision surveys like \textit{Euclid} and \textit{SKA}. The dependence on $\Delta$ arises uniquely from the Barrow entropy-induced modification to the agegraphic energy density, distinguishing this framework from generic quintom models that rely on multiple scalar degrees of freedom. Therefore, while the quintessence-to-phantom \textit{trend} is not exclusive to our models, the smooth asymptotic evolution toward phantom behavior—combined with model-specific controls such as $\Delta$—provides a distinctive signature of the EA-BADE and EA-NBADE reconstructions, even in the absence of a strict phantom crossing in $\omega_{DE}(z)$.

\begin{figure}[htbp]
    \centering
    \begin{subfigure}[b]{0.45\textwidth}
        \includegraphics[width=\textwidth]{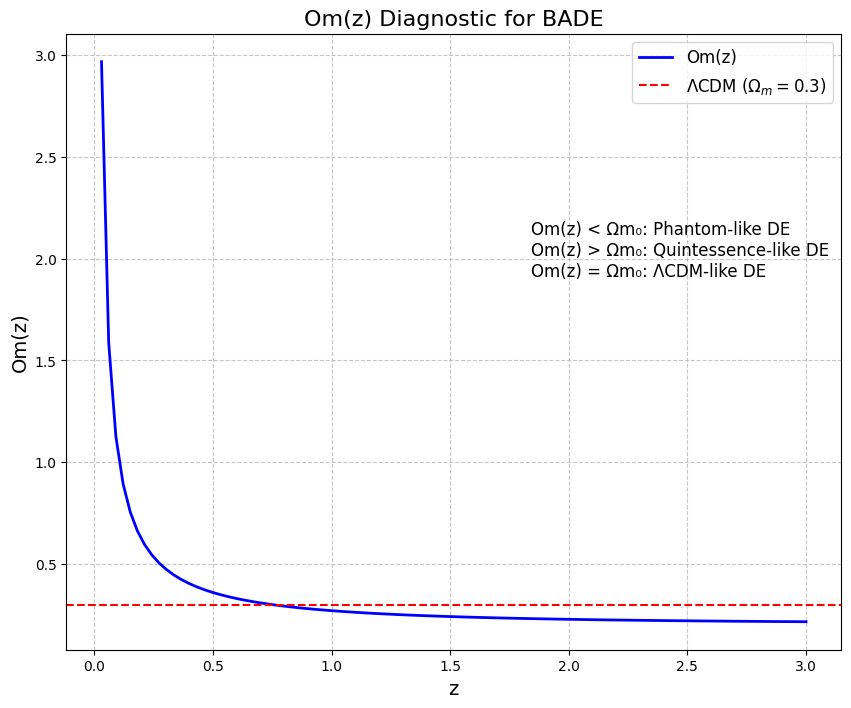}
        \caption{}
        \label{fig:om_BADE}
    \end{subfigure}
    \begin{subfigure}[b]{0.45\textwidth}
        \includegraphics[width=\textwidth]{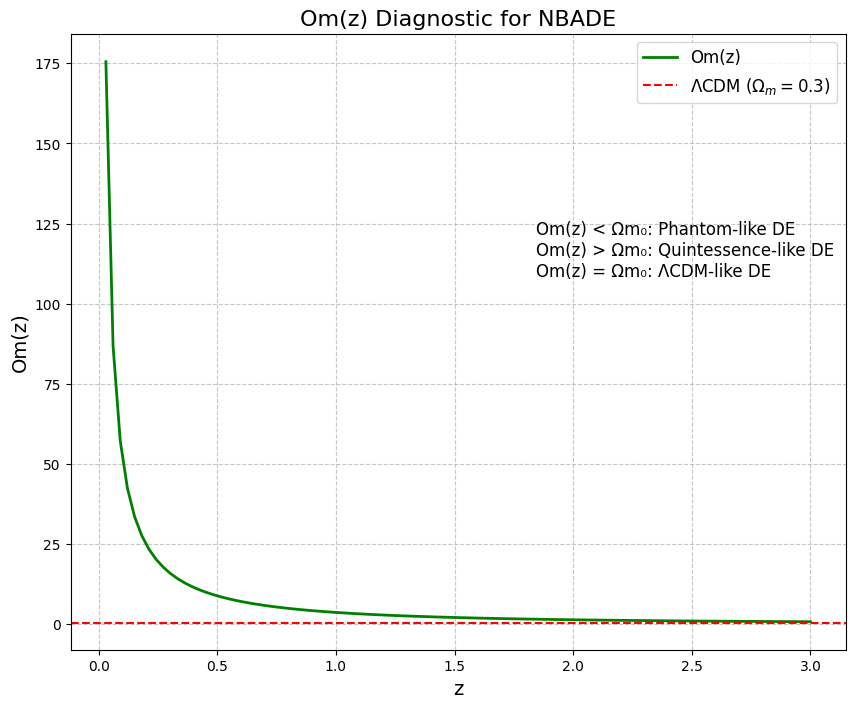}
        \caption{}
        \label{fig:om_NBADE}
    \end{subfigure}
    \caption{Plot of evolution trajectory of $Om(z)$ diagnostic parameter for the fig: (a) BADE model and fig: (b) NBADE model.}
    \label{fig:om_BADE_NBADE}
\end{figure}

{\bf Om(z) Diagnostic:} Figure~\ref{fig:om_BADE_NBADE} presents a comparative analysis utilizing the Om diagnostic parameter for both BADE and NBADE cosmological frameworks across varying redshifts. This diagnostic tool serves as a crucial discriminator for evaluating dark energy characteristics relative to the canonical $\Lambda$CDM paradigm. The evolutionary trajectories reveal distinct temporal behaviors. In the high-redshift domain [$(z > 0.5)$ for BADE and $(z > 1.0)$ for NBADE], the Om diagnostic demonstrates different characteristics: the BADE model maintains values marginally lower than the $\Lambda$CDM benchmark, while the NBADE framework shows remarkable alignment with the $\Lambda$CDM reference. 

A notable departure from $\Lambda$CDM behavior emerges at lower redshifts $(z < 0.6)$, characterized by a steep ascent in the $Om(z)$ function, which traverses the $\Lambda$CDM threshold and continues its upward trend. This phenomenon illuminates the dynamic nature of dark energy within these frameworks:

\begin{itemize}
    \item The low-redshift regime, characterized by $Om(z) > \Omega_m$, indicates quintessence-like properties in both BADE and NBADE scenarios.
    \item High-redshift analysis reveals contrasting behaviors: BADE exhibits phantom-like characteristics ($Om(z) < \Omega_m$), while NBADE maintains $\Lambda$CDM-consistent properties. The intermediate redshift range demonstrates a pronounced transition toward quintessence-like behavior, evidenced by the sharp increase in $Om(z)$.
\end{itemize}

These findings suggest a temporally evolving dark energy component, which exhibits increasing dominance in recent cosmic epochs. Such behavior potentially provides new perspectives on cosmic acceleration and the time-dependent properties of dark energy.

{\bf Other Cosmological and Model Parameters:} An examination of Table~\ref{tab: best_fit_params_values} reveals the following parameter constraints from observational data analysis:

The present-epoch matter density parameter $\Omega_{m0}$ demonstrates values of $0.314_{-0.066}^{+0.034}$ (BADE) and $0.309_{-0.002}^{+0.001}$ (NBADE) when constrained using Pantheon+SH0ES data. Analysis with CC+BAO yields $0.326_{-0.060}^{+0.040}$ (BADE) and $0.272_{-0.064}^{+0.066}$ (NBADE). These findings align with contemporary Planck measurements \cite{aghanim2020planck}.

Analysis of the matter equation of state parameter $\omega_m$ indicates near-zero values across both frameworks. Specifically, the BADE framework yields $\omega_m = 0.139_{-0.025}^{+0.036}$ (Pantheon+SH0ES) and $0.005_{-0.029}^{+0.038}$ (CC+BAO). The NBADE framework produces $\omega_m = 0.132_{-0.014}^{+0.015}$ (Pantheon+SH0ES) and $\omega_m = -0.028_{-0.043}^{+0.042}$ (CC+BAO). These measurements suggest predominantly dust-like matter behavior, with potential exotic matter contributions within Einstein-Aether gravity. The Barrow exponent parameter $\Delta$ exhibits distinctive values: BADE framework shows $\Delta = 0.324_{-0.201}^{+0.322}$ (Pantheon+SH0ES) and $\Delta = 0.389_{-0.275}^{+0.275}$ (CC+BAO), while NBADE yields $\Delta = 0.467_{-0.286}^{+0.351}$ (Pantheon+SH0ES) and $\Delta = 0.451_{-0.243}^{+0.331}$ (CC+BAO). These measurements support a non-flat universe hypothesis with negative curvature \cite{adhikary2021barrow}, diverging from flat universe assumptions in Einstein gravity \cite{anagnostopoulos2020observational}.

The coupling parameter $M$ is determined to be $M = 0.771_{-0.166}^{+0.163}$ (BADE) and $M = 0.553_{-0.243}^{+0.270}$ (NBADE) using Pantheon+SH0ES data, while CC+BAO analysis provides $M = 0.714_{-0.136}^{+0.180}$ (BADE) and $M = 0.626_{-0.301}^{+0.273}$ (NBADE). These results indicate subtle variations in distance measurements compared to standard $\Lambda$CDM predictions. The cosmic age parameter $t_{\text{age}}$ is constrained to $14.00^{+0.19}_{-0.17}$ Gyr (BADE) and $14.23^{+0.17}_{-0.19}$ Gyr (NBADE). These determinations remain consistent across datasets and correspond well with current observational estimates of universal age~\cite{aghanim2020planck}.

{\bf Comparative Analysis and Distinctive Features of EA-BADE/NBADE Models:} To address the unique characteristics of our combined Einstein-Aether (EA) gravity with Barrow and non-Barrow entropic dark energy models, we present a comprehensive comparative analysis with existing modified gravity and dark energy frameworks. While several cosmological models can reproduce similar behaviors in individual diagnostic parameters, the combined EA-BADE/NBADE framework offers distinct observational signatures that differentiate it from other theoretical constructs. For instance, the quintessence-like behavior observed in our $\omega_{DE}$ analysis is common in scalar field models like quintessence \cite{Tsujikawa_2013} and k-essence \cite{armendariz2001essentials}. However, unlike these models, our framework exhibits a distinctive evolution pattern in the statefinder plane that transitions from quintessence-like behavior through the $\Lambda$CDM fixed point to Chaplygin gas-like characteristics. This evolutionary trajectory is markedly different from that of standard scalar field models, which typically remain confined to a specific region of the statefinder plane \cite{sahni2003statefinder}.

The partial stability indicated by our $v_s^2$ analysis is particularly noteworthy when compared to other modified gravity theories. While $f(R)$ gravity models often demonstrate consistent stability across all redshift ranges \cite{nojiri2007introduction}, our EA-BADE/NBADE models exhibit a redshift-dependent stability profile with a transition at $z \approx 1$. This characteristic stability transition could serve as a distinctive observational signature for discriminating our models from other modified gravity frameworks. Furthermore, the behavior of the Om$(z)$ diagnostic reveals another unique aspect of our models. The sharp transition in the BADE model from phantom-like behavior at high redshifts to quintessence-like characteristics at low redshifts differs significantly from the behavior observed in scalar-tensor theories \cite{Boisseau_2000} and Galileon models \cite{Deffayet_2009}, which typically exhibit more gradual transitions or consistent behavior across redshift ranges. This distinct feature provides a potential observational test for the EA-BADE model, as future high-precision measurements of the expansion history could detect this characteristic transition. The cosmic age estimates derived from our models ($14.00^{+0.19}_{-0.17}$ Gyr for BADE and $14.23^{+0.17}_{-0.19}$ Gyr for NBADE) align well with current observational constraints while emerging from fundamentally different physical mechanisms than standard $\Lambda$CDM cosmology. It is important to acknowledge that certain individual features observed in our models can be reproduced by other theoretical frameworks. For example:

\begin{enumerate}
    \item {\bf Deceleration Parameter Evolution:} The transition redshift values we obtain ($z_t \approx 0.7568$ for BADE and $z_t \approx 0.6561$ for NBADE) fall within the range predicted by various modified gravity theories, including $f(T)$ gravity \cite{wu2010f} and Brans-Dicke theory \cite{banerjee2001brans}. However, the specific combination of transition redshift and present-day deceleration parameter values ($q_0 = -0.4985$ for BADE and $q_0 = -0.2759$ for NBADE) represents a distinctive signature of our models.

    \item {\bf Jerk, Snap, and Lerk Parameters:} The positivity of the jerk parameter across all redshifts is a feature shared with many dark energy models. However, the specific behavior of the snap parameter remaining consistently positive in both models, while the lerk parameter exhibits sign changes in the NBADE model, creates a unique cosmographic fingerprint that distinguishes our framework from other alternative gravity theories such as DGP \cite{dvali_DGP} and Gauss-Bonnet gravity \cite{Nojiri_2005_Another}.

    \item {\bf Equation of State Parameter:} The present-day values we obtain ($\omega_{DE0} = -0.8718$ for BADE and $\omega_{DE0} = -0.7841$ for NBADE) are within the quintessence regime, similar to many dynamical dark energy models. However, the particular evolutionary trajectory of $\omega_{DE}(z)$ in our models, especially when considered alongside the $\omega'_{DE}-\omega_{DE}$ phase plane analysis, exhibits distinctive features. Unlike standard quintessence models that typically show a thawing or freezing behavior \cite{Caldwell_2005}, our models demonstrate a more complex evolutionary pattern in the phase plane.

    \item {\bf Density Parameter Evolution:} The matter-dark energy equality redshift of $z \approx 0.5$ is consistent with standard cosmological models. However, when analyzed in conjunction with our $Om(z)$ diagnostic, which shows model-specific deviations from $\Lambda$CDM at different redshift ranges, our framework presents a unique cosmological signature.

    \item {\bf Statefinder Diagnostics:} While individual segments of the statefinder evolutionary trajectories can be reproduced by various modified gravity and dark energy models, the complete path through the $(r,s^*)$ and $(r,q)$ planes—beginning in the quintessence region, transitioning through the $\Lambda$CDM point, and ending in the Chaplygin gas region—represents a distinctive feature of our EA-BADE/NBADE framework. This particular evolutionary sequence is not typically observed in either pure modified gravity models like $f(R)$ theories \cite{Cao_2018} or standard dark energy models like phantom fields \cite{mhamdi2023comparing}.
\end{enumerate}

A key aspect that truly distinguishes our approach is the specific coupling mechanism between Einstein-Aether gravity and entropic dark energy formulations. This coupling, characterized by the parameter $M$ (measured as $M = 0.771_{-0.166}^{+0.163}$ for BADE and $M = 0.553_{-0.243}^{+0.270}$ for NBADE using Pantheon+SH0ES data), introduces unique effects on cosmological evolution that cannot be readily replicated by either pure modified gravity or pure dark energy models. The Barrow exponent parameter $\Delta$ in our models ($\Delta = 0.324_{-0.201}^{+0.322}$ for BADE and $\Delta = 0.467_{-0.286}^{+0.351}$ for NBADE using Pantheon+SH0ES data) introduces an additional layer of complexity that distinguishes our framework from conventional approaches. This parameter, arising from considerations of modified black hole entropy due to quantum-gravitational effects, introduces a fundamentally different physical mechanism for cosmic acceleration compared to standard scalar field or modified gravity approaches.

\begin{table}[h]
\centering
\small
\caption{Comparative cosmological diagnostics for dark energy models}
\label{tab:comparative_diagnostics_summary}
\renewcommand{\arraystretch}{1.5} % Vertical stretch
\begin{tabular}{|l|p{2.8cm}|p{2.2cm}|p{2.8cm}|p{2.8cm}|c|}
\hline
\textbf{Model} & \textbf{$\omega_{DE}$ Behavior} & \textbf{Ph. Cross} & \textbf{Om$(z)$ Shape} & \textbf{$(r,s)$ Trend} & \textbf{Ref.} \\
\hline
EA-BADE & $-1 < \omega < -0.7 \to \omega < -1$ & Yes ($z \approx -0.5$) & Concave up ($z<1$), down ($z>1$) & $(1,1)\to(1,0)$ (de Sitter) & -- \\
\hline
EA-NBADE & $-1 < \omega < -0.8 \to \omega < -1$ & Yes ($z \approx -0.6$) & Flat ($z<0.5$), incr. ($z>0.5$) & $(1,1)\to(1,0)$ (de Sitter) & -- \\
\hline
Quintessence & $-1 < \omega < -0.3$ & No & Concave down (all $z$) & $(r<1, s>0) \to (1,0)$ & \cite{RevModPhys.75.559} \\
\hline
Phantom & $\omega < -1$ & No & Concave up (all $z$) & $(r>1, s<0) \to (1,0)$ & \cite{Caldwell_2002} \\
\hline
f(R) Gravity & $-1.2 < \omega_{\text{eff}} < -0.8$ & Possible & Convex/concave & Smooth to $(1,0)$ & \cite{de2010f} \\
\hline
Tsallis HDE & $\omega \to <-1$ ($\delta$ tuned) & Yes ($\delta>1$) & Curved ($\delta$-dep.) & Distinct ($r \neq 1, s \neq 0$) & \cite{Saridakis_2018} \\
\hline
Chaplygin Gas & $\omega \to -1$ (late) & No & Flat (late) & $(1,1) \to (1,0)$ & \cite{Bento_2002} \\
\hline
\end{tabular}

\end{table}

When these distinctive features are considered collectively rather than individually, our EA-BADE/NBADE framework emerges as a novel theoretical construct with unique observational signatures. Future high-precision cosmological observations, particularly those targeting the expansion history and growth of structure at intermediate redshifts ($0.5 < z < 2.0$), would be particularly valuable for testing the distinctive predictions of our models against other competing theoretical frameworks. In conclusion, while individual cosmological parameters in our models may exhibit behaviors similar to those in other theoretical frameworks, the specific combination of parameter values, evolutionary trajectories, and underlying physical mechanisms in the EA-BADE/NBADE models creates a distinctive cosmological signature that can be tested against future observational data. This holistic approach to model discrimination, rather than relying on individual diagnostic parameters, provides a more robust framework for evaluating the viability of our theoretical constructs against the broader landscape of cosmological models.

\section{Conclusions} \label{Sect: Conclusion}

In this work, we performed a comprehensive reconstruction of Einstein-Aether (EA) gravity by integrating two modified agegraphic dark energy models—the Barrow Agegraphic Dark Energy and its extension, the New Barrow Agegraphic Dark Energy. The reconstruction framework involved equating the energy densities of BADE and NBADE with that of EA gravity to derive the form of the function $F(K)$, which characterizes the aether kinetic term. This approach provides a novel entropy-motivated perspective to explore dynamical dark energy in the Einstein-Aether context. We showed that the reconstructed $F(K)$ functions exhibit distinct monotonic behaviors depending on the model and parameter choices, particularly the Barrow exponent $\Delta$. While BADE produces an increasing $F(K)$ with respect to $K$, NBADE yields a decreasing profile, reflecting differing dynamical behavior rooted in the entropy assumptions.

Using a robust statistical analysis based on Markov Chain Monte Carlo (MCMC) techniques and observational constraints from CC+BAO and Pantheon+SH0ES data, we constrained the key model parameters, including $\Omega_{m0}$, $\omega_m$, $\Delta$, $M$, and $t_{\text{age}}$. The models produce Hubble parameter evolution in good agreement with observational measurements, though mild tensions with SH0ES remain—similar to what is observed in other non-$\Lambda$CDM models. The transition from decelerated to accelerated expansion is observed near $z_t \approx 0.7568$ (BADE) and $z_t \approx 0.6561$ (NBADE), consistent with constraints from SNIa, BAO, and CMB data. Higher-order kinematic indicators—jerk, snap, and lerk—suggest a dynamic acceleration history, and the statefinder diagnostics $(r,s^*)$ and $(r,q)$ place the models near quintessence and Chaplygin-like regimes, asymptotically approaching de Sitter behavior.

The dark energy equation of state $\omega_{DE}$ remains negative across the evolution history, indicative of persistent acceleration. Both models exhibit a current quintessence-like behavior transitioning toward a phantom regime in the far future. This quintessence-to-phantom evolution is also captured in the $\omega_{DE}'$–$\omega_{DE}$ phase space and Om(z) diagnostics. Although this feature is common among various dynamical dark energy models, the redshift of transition and the corresponding phase space trajectories in our EA-BADE and EA-NBADE frameworks may provide distinguishing signatures when evaluated against future high-precision datasets. Stability analysis using the squared sound speed $v_s^2$ reveals partial epoch-dependent stability, with the models being stable at higher redshifts and exhibiting instabilities near the current epoch. This behavior is consistent with some other noncanonical DE models and signals a need for further theoretical scrutiny regarding perturbative behavior in modified gravity scenarios.

In comparative context, while many of the observed features—such as late-time acceleration, freezing behavior in the $\omega_{DE}'$–$\omega_{DE}$ plane, and redshift-dependent Om diagnostics—are shared by other well-studied models (e.g., CPL, Holographic, or quintom-type models), the EA-BADE and EA-NBADE reconstructions offer a distinct pathway by incorporating entropy-based corrections to agegraphic energy density. The Barrow exponent $\Delta$, in particular, plays a critical role in modulating model behavior and could serve as a key discriminator in future observational tests.

Although the models do not yet demonstrate a decisive improvement over $\Lambda$CDM in terms of statistical significance, their ability to encode entropy deformation in a covariant gravitational framework provides a fertile ground for further theoretical developments. The models serve as useful phenomenological laboratories to probe the interface of modified gravity, nonextensive thermodynamics, and cosmic acceleration. Future research should examine possible time variation of the Barrow exponent $\Delta$—as suggested by recent studies—and explore generalized entropic frameworks such as the Nojiri-Odintsov-Paul entropy formalism. Additionally, incorporating interaction terms between dark energy and dark matter, as well as leveraging more extensive datasets (e.g., CMB, LSS, and gravitational wave backgrounds), could sharpen parameter constraints and assess the long-term viability of the models. Finally, implementing these models in large-scale structure simulations would allow for direct comparison with $\Lambda$CDM predictions and further illuminate the potential of Einstein-Aether gravity in the context of cosmic evolution. In summary, while not conclusively superior to standard cosmological models, the EA-BADE and EA-NBADE frameworks represent a promising and theoretically motivated extension of Einstein-Aether gravity. Their compatibility with observational data and capacity to encapsulate non-standard thermodynamic corrections position them as viable candidates for further exploration in the quest to understand the nature of dark energy and the fundamental structure of spacetime.

\section*{Acknowledgments}
A.K. is thankful to IIEST, Shibpur, India for Institute Fellowship
(JRF).

\section*{Data Availability Statement}
The data generated or analyzed in this study are contained within the published article. No new data were created or examined during this research.

\section*{ Conflict of Interest Statement}
The authors state that this research was conducted without any commercial or financial relationships that could be seen as a potential conflict of interest.

% After the Acknowledgments section
\section*{Author Contributions}
% Details of contributions
Conceptualization, B.C. and T.M.; Methodology, T.M. and B.C.; Software, T.M. and B.C.; Formal Analysis, B.C., T.M. and A.K.; Writing—Original Draft Preparation, B.C. and T.M.; Writing—Review and Editing, A.K. and U.D.; Visualization, T.M. and B.C.; Supervision, U.D. All authors have read and agreed to the published version of the manuscript.

% References
\bibliographystyle{elsarticle-num}  % Elsevier style bibliography
\bibliography{references}           % Name of the .bib file (without the extension)

\end{document}